\documentclass[epj]{webofc}
\usepackage[varg]{txfonts}   
\wocname{EPJ Web of Conferences}
\woctitle{ICNFP 2015}
\begin{document}
\selectlanguage{english}
\title{Highlights from the COMPASS experiment at CERN 
}
%
%
\subtitle{Hadron spectroscopy and excitations}

\author{Frank Nerling\inst{1}\fnsep\thanks{\email{nerling@cern.ch}}, on behalf of the COMPASS Collaboration 
}

\institute{Universit\"at Mainz, Institut f\"ur Kernphysik, Germany 
}

\abstract{%
The COMPASS experiment at the CERN-SPS studies the spectrum and the 
structure of hadrons by scattering high energy hadrons and polarised
muons off various fixed targets. Recent results for the hadron programme 
comprise highlights from different topics. A selective overview is given 
and, among others, the following results are discussed. 
The precise determination of the pion polarisability, a long standing 
puzzle that has been solved now, is presented as well as measurements 
of radiative widths. The observation of a new narrow axial-vector state, 
the $a_1(1420)$, as well as deeper insights into the exotic $1^{-+}$-wave, 
which is under study since decades by several experiments, are discussed 
and further, the search for the charmonium-like exotic $Z_c(3900)$ state 
in the COMPASS data is covered.
}
\maketitle
\section{Introduction}
\label{intro}
The COMPASS fixed-target experiment~\cite{compass} at CERN-SPS is a facility to study quantum chromodynamics (QCD).
It is dedicated to study the perturbative and non-perturbative regime of QCD, and probe the structure and dynamics of hadrons.
Using a muon beam, COMPASS investigates nucleon structure by measuring helicity, transversity 
and general parton distribution functions --- the recent highlights from the COMPASS muon programme 
are summarised in~\cite{platchkov}.  
Hadron excitations and spectroscopy, including the search for exotic states, are investigated using hadron beams on a 
liquid hydrogen (proton) and different nuclear targets. A selection of recent results of the COMPASS hadron programme 
are discussed and summarised here.
 
\begin{figure}[tp!]
    \begin{center}
      \vspace{-0.5cm}
\resizebox{1.0\columnwidth}{!}{%
  \includegraphics[clip,trim= 15 70 5 100, width=1.0\linewidth, angle=0]{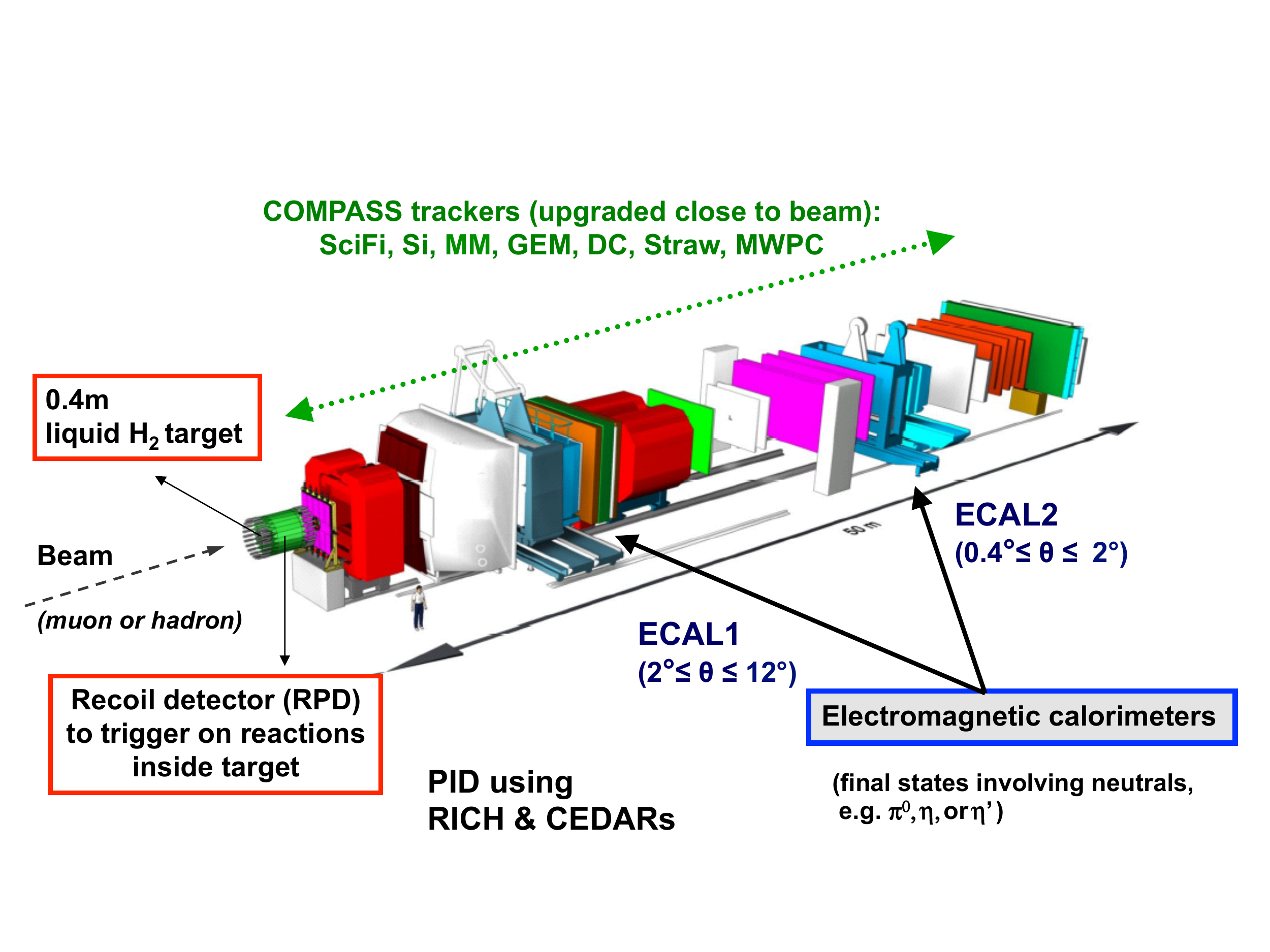} }
    \end{center}
    \begin{center}
     \vspace{-0.3cm}
     \caption{Schematic drawing of the COMPASS two-stage spectrometer as used for the hadron runs 2008/09.}  
        \label{fig:CompassApp}
     \end{center}
     \vspace{-0.8cm}
\end{figure}

The two-stage spectrometer (Fig.\,\ref{fig:CompassApp}) is equipped with electromagnetic and hadronic calorimeters, providing 
detection of charged and neutral final state particles with a homogeneous acceptance over a wide 
kinematic range, especially covering a large range in momentum transfer. 
At lowest reduced four-momentum transfers between the beam particle and the target of $t'$\,$<$\,$10^{-3}$\,(GeV/$c$)$^2$, Primakoff 
reactions are studied, from which the pion polarisability has been extracted at high experimental precision, Chiral perturbation 
theory (ChPT) is tested and the radiative decay widths of the $a_2(1320)$ and the $\pi_2(1670)$ have been 
measured in pion-photon reactions. At low four-momentum transfers of 0.1\,(GeV/$c$)$^2$\,$<$\,$t'$\,$<$\,1\,(GeV/$c$)$^2$, the mass 
spectrum of hadrons is investigated, including searches for (spin-) exotic mesons.  

\section{Results on hadron excitations}
\label{sec-1}
For the understanding of QCD and the low-momentum expansion ChPT, in which the pions are identified as the 
Goldstone bosons as a result of the spontaneous chiral symmetry breaking, the pion properties are a crucial input. 
Whereas pion-pion scattering has extensively been studied and successfully described within ChPT, experimental 
results of pion-photon reactions in terms of the most basic process provided still a puzzle:
Previous measurements of the pion polarisability, the leading structure-dependent term of Compton scattering, 
provided resultant values significantly larger as expected from theory, {\it cf.}\,Fig.\,\ref{fig:piPol}. Apart from the pion 
polarisability in Primakoff reactions (Sec.\,\ref{subsec-piPol}), also pion-photon interactions with more than 
one pion in the final state are studied (Secs.\,\ref{subsec-chiralDynamics} and \ref{subsec-radWidths}).         

\subsection{Measurement of pion polarisability}
\label{subsec-piPol}
The rigidity of an extended object against deformations by an external electric 
or magnetic field are described by the so-called electric and magnetic polarisabilities
$\alpha$ and $\beta$, respectively. As the pion is not a point-like particle, it is 
sensitive to deformations by such external fields, and this tiny effect, called pion polarisability,
is under the assumption of $\alpha_\pi =-\beta_\pi$ predicted in Chiral Perturbation Theory 
(ChPT) to be $5.7\times 10^{-4}$\,fm$^3$~\cite{pionPolPredict}.

Polarisabilities are usually measured in Compton scattering experiments. Even though 
pions can hardly be used as a fixed-target, they can be scattered off the Coulomb 
potential of a heavy nucleus, like a Nickel (Ni) target. In the corresponding 
$\pi^-\gamma \to \pi^-\gamma$ scattering process, the polarisabilities manifest as a 
deviation from the Born cross-section of a point-like particle. 
The measurement of the cross-section of this process allows to extract the pion 
polarisabilities via the deviation from the assumption for a point-like particle. 

Such a measurement is experimentally demanding and the systematics need precisely to 
be controlled. An important advantage of the COMPASS experiment is that we can perform 
this measurement with the pion and with the muon beam (quickly switchable within about 
two hours), in which the latter acts as a control measurement of the point-like muon. 
 
\begin{figure}[tp!]
  \begin{minipage}[h]{.65\textwidth}
    \begin{center}
      \vspace{-0.5cm}
\resizebox{1.0\columnwidth}{!}{%
     \includegraphics[clip,trim= 5 5 5 0, width=0.87\linewidth, angle=0]{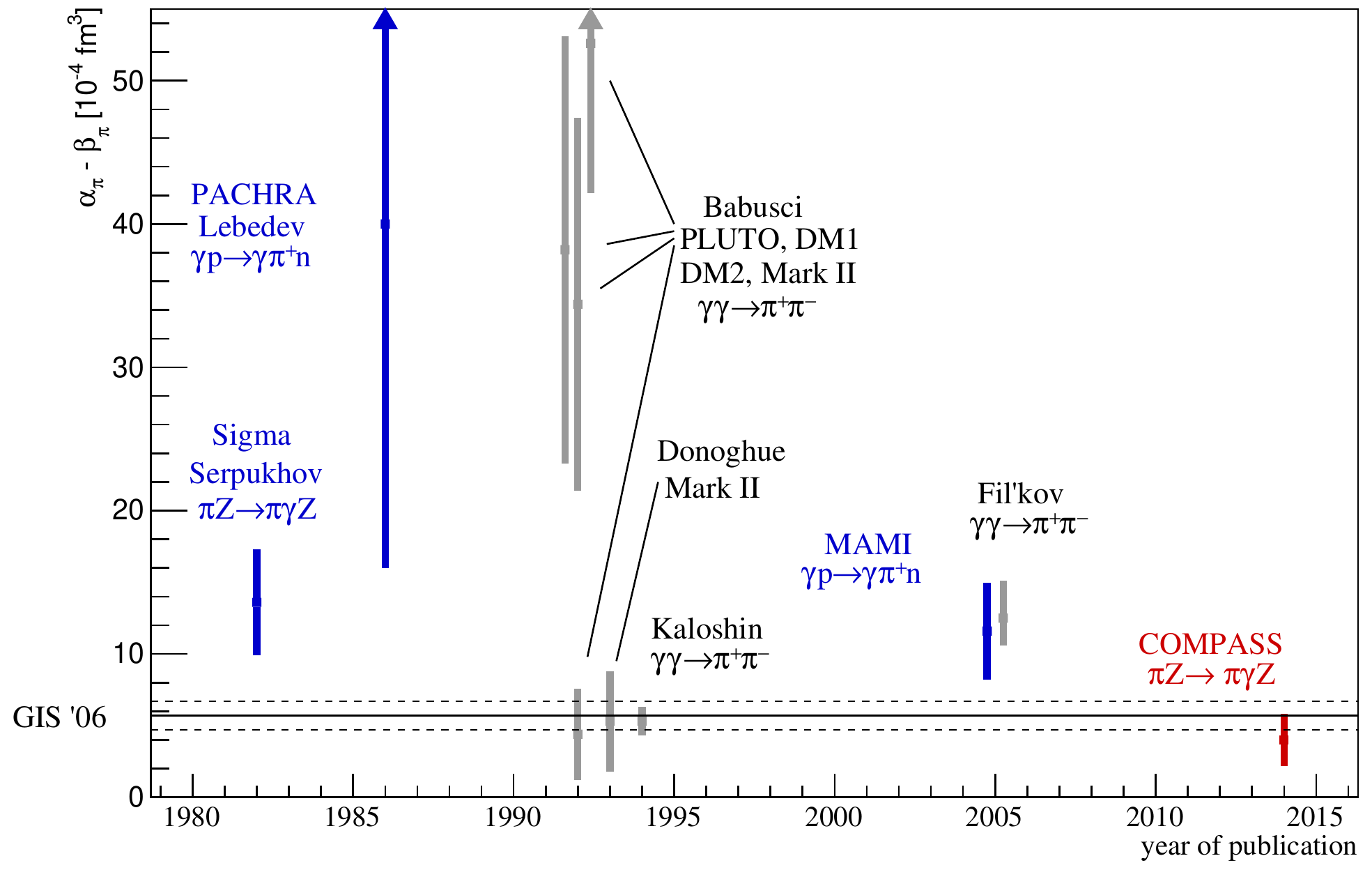} }
    \end{center}
  \end{minipage}
  \hfill
  \begin{minipage}[h]{.35\textwidth}
    \begin{center}
      \vspace{-0.5cm}
\resizebox{1.0\columnwidth}{!}{%
  \includegraphics[clip,trim= 15 5 5 18, width=1.0\linewidth, angle=0]{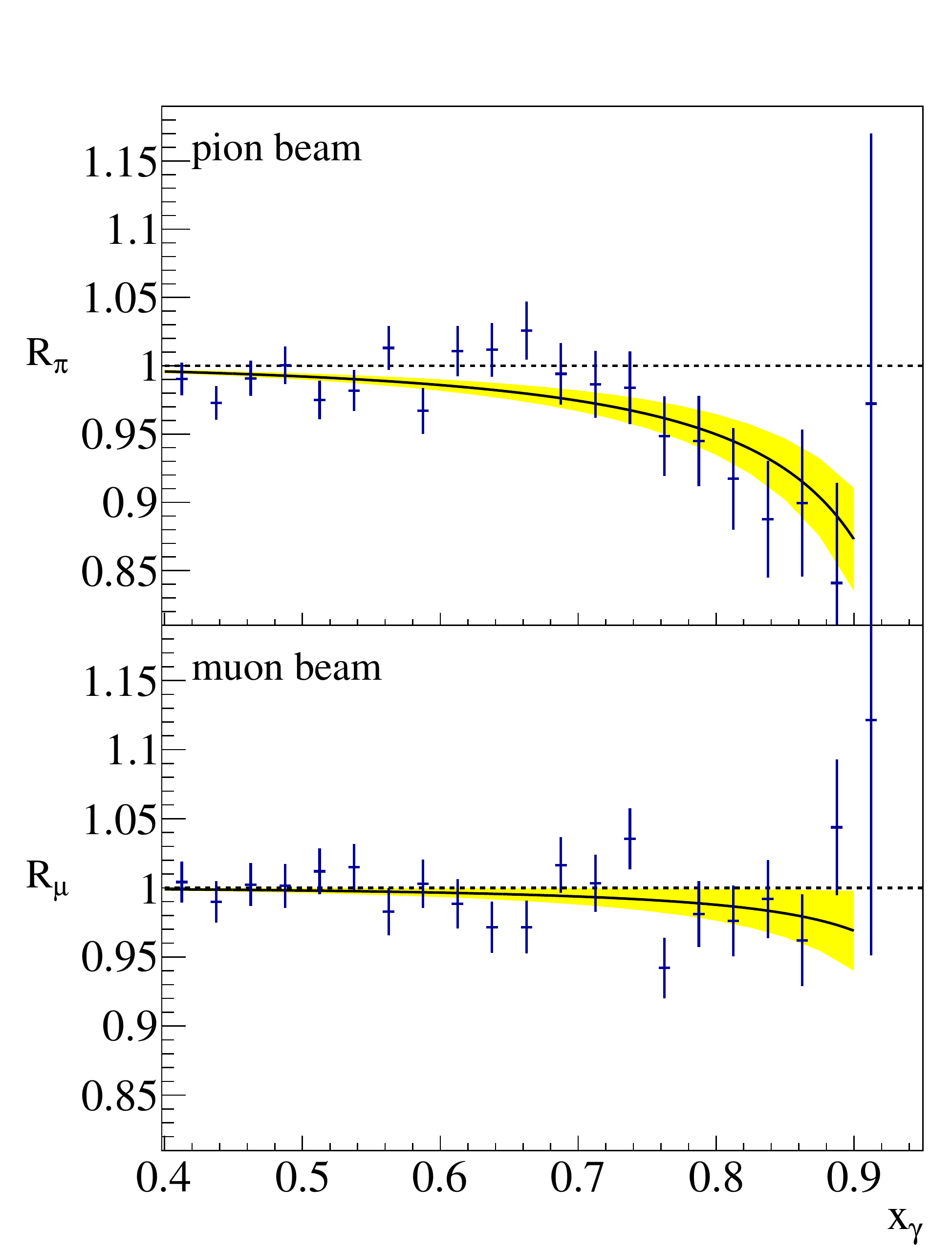} }
    \end{center}
  \end{minipage}
    \begin{center}
     \vspace{-0.3cm}
     \caption{{\it Left}: Compilation of experimental measurements of the pion polarisability $\alpha_\pi -\beta_\pi$ 
         including the new COMPASS result and the theoretical prediction from ChPT~\cite{pionPolPredict}, plot taken 
         from~\cite{nagelPhD}. 
         {\it Right:} Ratio $R$ of the measured cross-section over the simulated one for a point-like object for 
         the pion beam data {\it (top)} and for the control measurement with the muon beam data {\it (bottom)}~\cite{compassPiPol}.}  
        \label{fig:piPol}
     \end{center}
     \vspace{-0.8cm}
\end{figure}

The COMPASS data of scattering a pion beam off a Ni target comprises the Compton 
part of the cross-section~\cite{compassPiPol}. The $\pi^-\,Ni \to \pi^-\gamma \, Ni$ events 
are selected by requiring one negatively charged track from the vertex inside the Ni target, 
a high-energetic shower in the electromagnetic calorimeters for the final state photon, 
whereas the photon exchange is ensured by a cut on low photon virtualities $Q^2$,   
energy conservation is applied to select exclusive reactions, 
{\it i.e.} $\Delta E = E_{beam}- E_{\pi'} - E_\gamma \approx 0$, neglecting the (tiny) nuclear 
recoil energy. In order to stay in the kinematic region of stable trigger and muon 
identification efficiencies, the ratio of the final state photon and the initial beam 
energy $x_\gamma = E_{\gamma}/E_{beam}$ is required to be between 0.4 and 0.9. 
Under the assumption of $\alpha_\pi = -\beta_\pi$ that has to be made due to the limited 
statistics at lower $x_\gamma$ ({\it i.e.} the phase space region sensitive to $\alpha_\pi + \beta_\pi$),
the polarisation should manifest in terms of a decreasing ratio $R$ of the measured 
cross-section over the simulated one for a point-like particle with increasing $x_\gamma$.
This behaviour has exactly been measured for the pion beam data, {\it i.e.} for the extended 
object of the pion, as displayed in Fig.\,\ref{fig:piPol} (right, top). 
To extract the polarisability $\alpha_\pi$ from the measured ratio $R$, the photon energy spectrum
is examined according to the following relation, which is a simplified relation of the Primakoff 
cross-section in one-photon exchange approximation:
\begin{equation} \nonumber
R = \frac{\sigma(x_\gamma)}{\sigma_{\alpha_\pi=0}(x_\gamma)} = 1 -\frac{3}{2} \cdot \frac{{m_{\pi}}^3}{\alpha} \cdot \frac{ {x_\gamma}^2}{1-x_\gamma} \alpha_\pi.
\end{equation}
For the muon beam data (Fig.\,\ref{fig:piPol}, right, bottom), the size of the fake polarisability 
of $(0.5 \pm 0.5_{stat}) \times 10^{-4}$\,fm$^3$ is within the statistical uncertainty compatible with zero. 
This value is taken as an estimate of the systematic error due to apparative imperfections, not described 
by the MC simulation. For the pion beam data (Fig.\,\ref{fig:piPol}, right, top), a pion polarisability of 
$\alpha_\pi = (2.0 \pm 0.6_{stat})\times 10^{-4}$\,fm$^3$ is determined from the fit.
This pion polarisability value measured by COMPASS is in tension with previous experiments and in 
agreement with the prediction from ChPT (Fig.\,\ref{fig:piPol}, left).
\subsection{Measurement of chiral dynamics}
\label{subsec-chiralDynamics}
From the same data, further processes on chiral dynamics are accessible, like {\it e.g.} the chiral 
anomaly in $\pi^-\gamma \to \pi^-\gamma$, the corresponding analysis is still under way, the 
same holds for the neutral pion case of two-pion production $\pi^-\gamma \to \pi^-\pi^0\pi^0$ 
at low energy. Based on the 2004 pilot run data (Primakoff kinematics), the two-pion 
production for the charged case $\pi^-\gamma \to \pi^-\pi^+\pi^-$ has been 
analysed~\cite{compass_chiralDynamics}.

\begin{figure}[tp!]
  \begin{minipage}[h]{.48\textwidth}
    \begin{center}
      \vspace{-0.5cm}
\resizebox{1.0\columnwidth}{!}{%
  \includegraphics[clip,trim= 5 5 5 5, width=1.0\linewidth, angle=0]{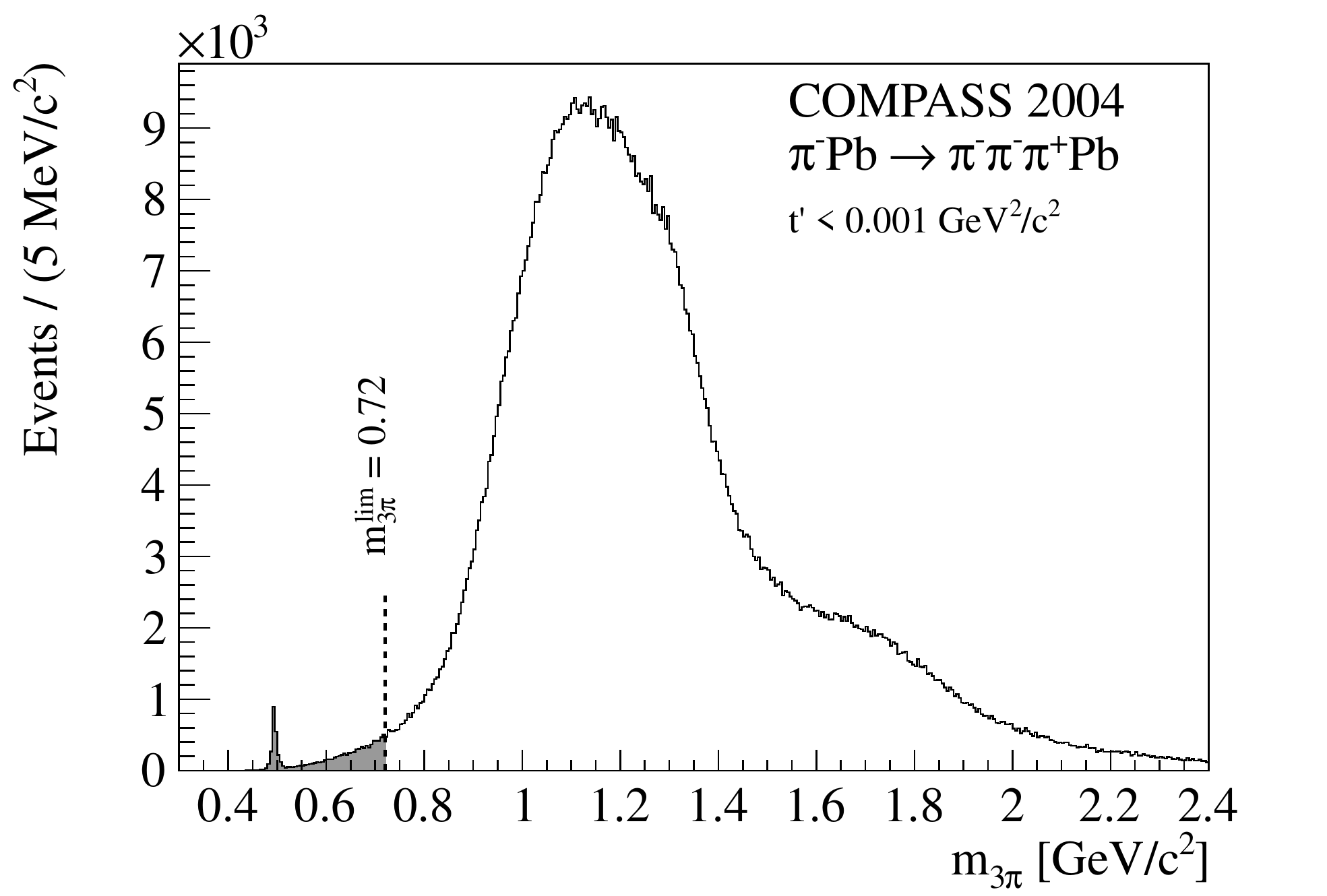} }
    \end{center}
  \end{minipage}
  \hfill
  \begin{minipage}[h]{.48\textwidth}
    \begin{center}
      \vspace{-0.5cm}
\resizebox{1.0\columnwidth}{!}{%
     \includegraphics[clip,trim= 5 5 5 5, width=1.0\linewidth, angle=0]{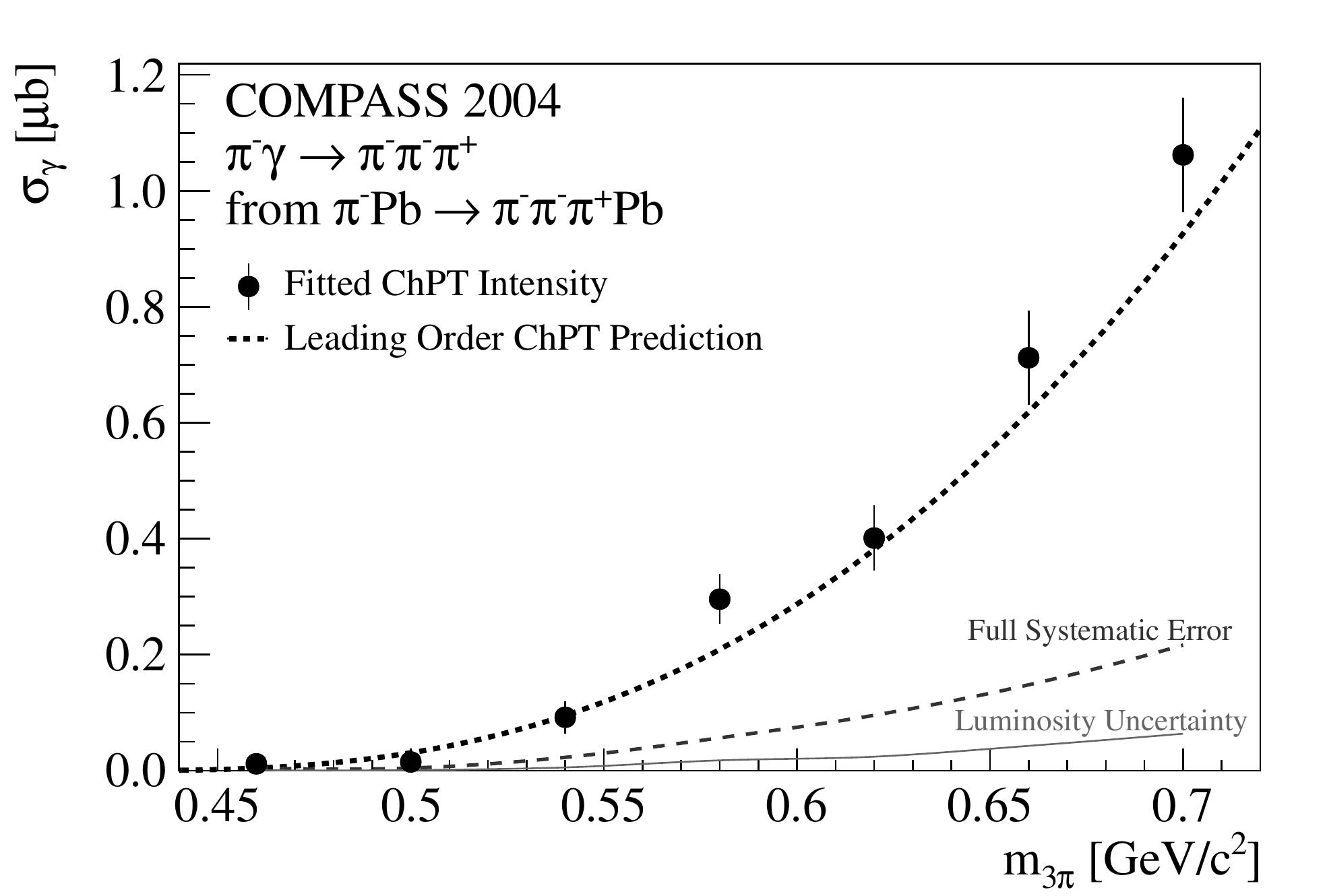} }
    \end{center}
  \end{minipage}
    \begin{center}
     \vspace{-0.3cm}
     \caption{{\it Left:} Spectrum of the $3\pi$ mass from $\pi^- Pb \to \pi^-\pi^+\pi^- Pb$ in the Primakoff region of 
       four-momentum transfer $t' < 10^{-3}$\,GeV$^2$/$c^2$. The region of interest in view of ChPT up to 
       $m_{3\pi} \approx 0.72$\,GeV/$c^2$ is indicated. At $m_{K^-} \approx 0.49\,$GeV/$c^2$, a peak from the admixture of 
       incident beam kaons decaying into $(3\pi)^-$ is visible that are used for flux normalisation. 
       {\it Right:} First measurement of the photo-production cross-section $\sigma_\gamma$ in this mass range, found in 
       agreement with the prediction from leading order ChPT~\cite{compass_chiralDynamics}.
     }  
        \label{fig:chiralDynamics}
     \end{center}
     \vspace{-0.8cm}
\end{figure}

Figure~\ref{fig:chiralDynamics} (left) shows the mass spectrum of the outgoing three-pion final 
state. Of particular interest in view of chiral dynamics is the low mass tail up to three pion 
masses, as indicated in the figure. Performing a partial-wave analysis (PWA) including amplitudes from ChPT calculations substituting 
the isobaric partial waves in this mass range, after flux normalisation on the observed kaon 
decays (using the kaon admixture in the beam), the absolute cross-section is, within the 
experimental uncertainty of 20\,\%, found in agreement with the leading order ChPT 
prediction.   
     
In the $3\pi$ mass spectrum (Fig.\,\ref{fig:chiralDynamics}, left), structures from the well-known 
resonances $a_1(1260)$ and $a_2(1320)$ are visible at higher masses --- the corresponding radiative 
couplings have been measured as discussed in the next section (Sec.\,\ref{subsec-radWidths}).

\subsection{Measurement of radiative width}
\label{subsec-radWidths}
Electromagnetic transitions can be measured via radiative decays of a resonance $X \to \pi\gamma$.
The direct measurement of $\pi\gamma$ is experimentally difficult. Alternatively, the 
inverse process of scattering a pion off a Coulomb field producing an intermediate resonance $X$, 
{\it i.e.} the Primakoff reaction, can be used. The Coulomb potential of a heavy nucleus acts as a (quasi-real) 
photon source, and the Primakoff production cross-section of a resonance $X$ is proportional to the width 
of the radiative decay of that resonance: $\sigma_{\rm Primakoff} (X) \sim \Gamma_0(X \to \pi\gamma)$. 
Due to the exchange particle being a quasi-real photon, Primakoff produced resonances have predominantly 
spin projection $M$=$1$. Even though states with spin projection $M$=$1$ can also be produced diffractively, 
the cross-section here is proportional to $t'^{|M|}e^{-bt'}$, so that the diffractive production is highly 
suppressed at very small values of $t'$.  

\begin{figure}[tp!]
  \begin{minipage}[h]{.48\textwidth}
    \begin{center}
      \vspace{-0.5cm}
\resizebox{1.0\columnwidth}{!}{%
     \includegraphics[clip,trim= 5 5 5 5, width=1.0\linewidth, angle=0]{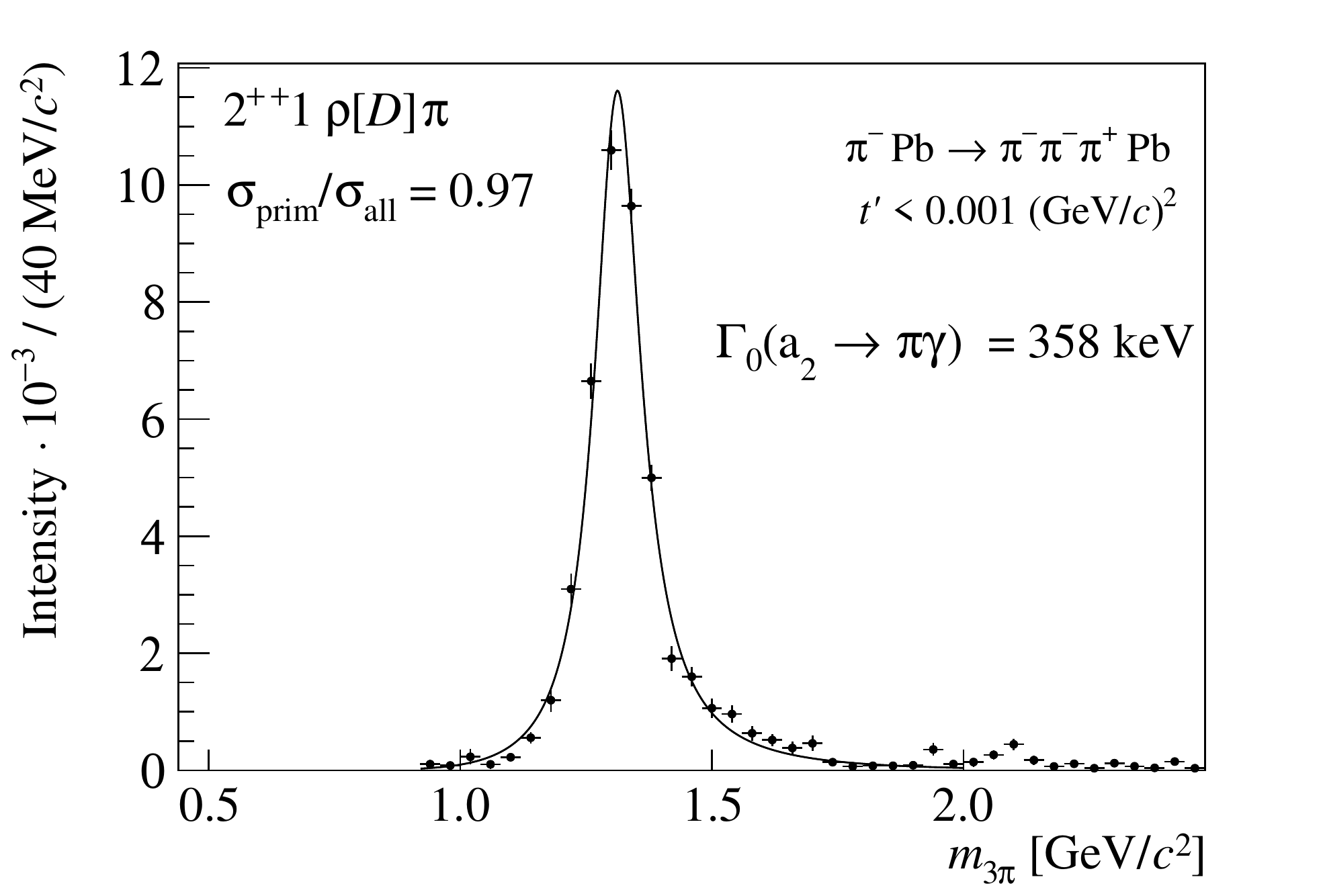} }
    \end{center}
  \end{minipage}
  \hfill
  \begin{minipage}[h]{.48\textwidth}
    \begin{center}
      \vspace{-0.5cm}
\resizebox{1.0\columnwidth}{!}{%
  \includegraphics[clip,trim= 5 5 5 5, width=1.0\linewidth, angle=0]{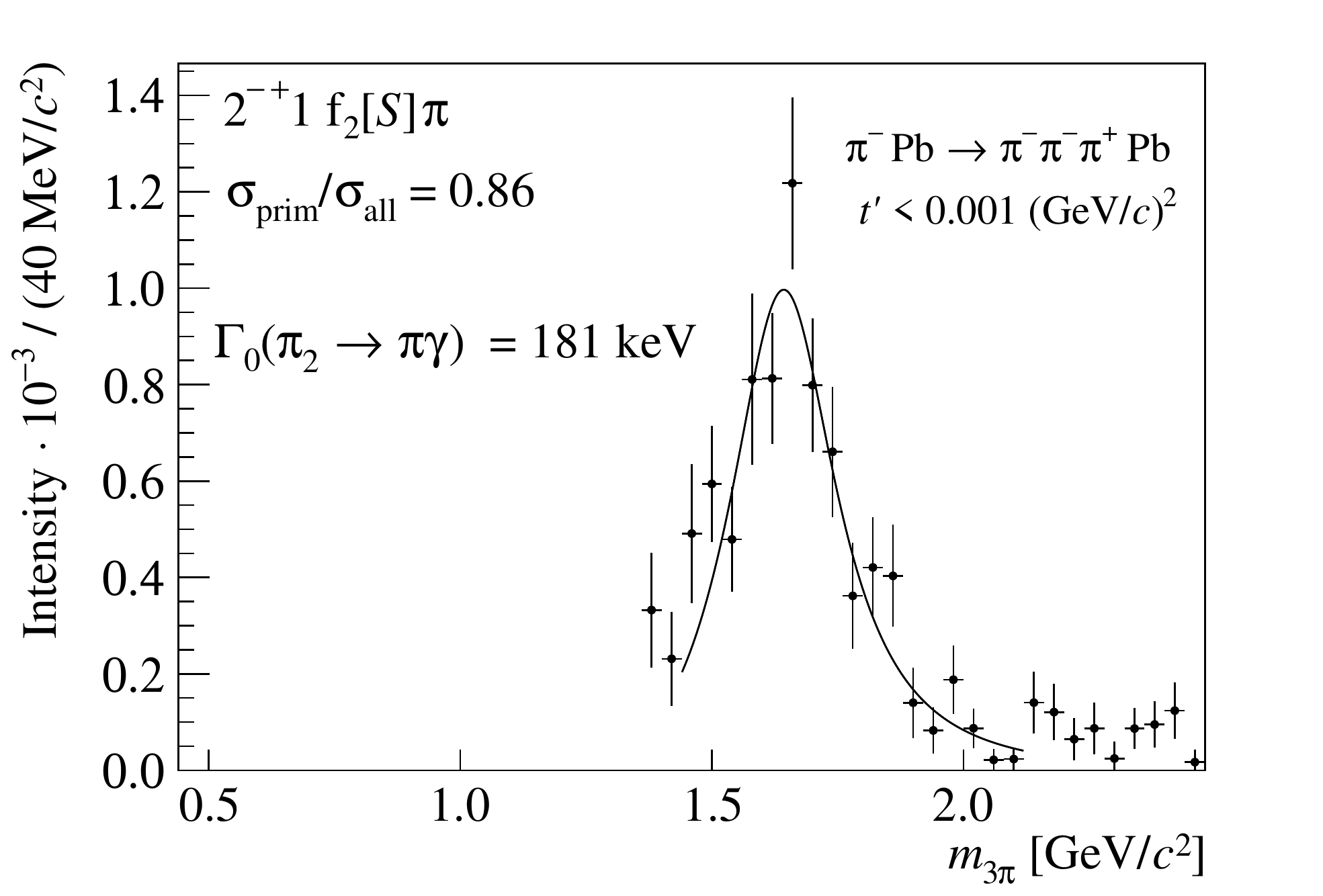} }
    \end{center}
  \end{minipage}
    \begin{center}
     \vspace{-0.3cm}
     \caption{Results of the measured radiative widths of the $a_2(1320)$ {\it (left)} and the 
       $\pi_2(1670)$ {\it (right)} (reprinted with kind permission of EPJ from~\cite{compass_radWidth}, 
       copyright Societ\`{a} Italiana di Fisica / Springer-Verlag 2014).
       As compared to previous measurements, 
       the COMPASS result for the $a_2(1320)$ is the most precise one, whereas for the $\pi_2 (1670)$,
       the COMPASS value is the first measurement at all.}  
        \label{fig:radWidth}
     \end{center}
     \vspace{-0.8cm}
\end{figure}

Based on the negative pion beam data on a thin lead (Pb) target, the COMPASS Collaboration studied this 
reaction and measured the radiative widths of the $a_2(1320)$ and the $\pi_2(1670)$~\cite{compass_radWidth}. 
Performing a partial-wave analysis, the data is decomposed into spin-parity states, and thus also the $M$=$1$ 
states are identified. 
To extract the radiative widths of the $a_2(1320)$ and the $\pi_2(1670)$, the two partial waves 
$J^{PC}M^\epsilon\,{\it (isobar)}\,\pi$ = $2^{++}1^\epsilon\rho(770)\pi$\,D-wave and $2^{-+}1^\epsilon f_2(1270)\pi$\,S-wave have been considered, respectively. 
For the events analysed, namely at $t'$<$10^{-3}\,$GeV$^2/$c$^2$, the contribution of Primakoff production 
is 97\,\% and 86\,\% of the intensities found in the $2^{++}1^\epsilon\rho(770)\pi\,D$ and 
$2^{-+}1^\epsilon f_2(1270)\pi$\,S-waves, respectively. The intensities for the two different 
reflectivities $\epsilon = \pm 1$ have to be summed up incoherently as the production plane at such small $t'$ 
is not measured sufficiently precise due to limited experimental resolution.
That there are further possible decay channels (than $\rho(770)\pi$ for the $a_2(1320)$ and $f_2(1270)\pi$ for 
the $\pi_2(1670)$) has been taken into account by consideration of the relevant branching fractions.
The fits to the data are shown in Fig.\,\ref{fig:radWidth}, from which a width of 
$\Gamma_0(a_2(1320)$\,$\to$\,$\pi\gamma)$\,$=$\,(358$\pm$6$\pm$42)\,keV and 
$\Gamma_0(\pi_2(1670)$\,$\to$\,$\pi\gamma)$\,$=$\,(181$\pm$11$\pm$27)\,keV have been measured for the 
$a_2(1320)$ and the $\pi_2(1670)$, respectively. Both results are consistent with predictions based on 
the VMD model~\cite{VMDmodel}. Compared to previous measurements, the COMPASS result for $\Gamma_0(a_2(1320)$\,$\to$\,$\pi\gamma)$ 
is the most precise one, whereas for $\Gamma_0(\pi_2(1670)$\,$\to$\,$\pi\gamma)$ the COMPASS result is the first measurement at all.

\section{Results on hadron spectroscopy}
\label{sec-2}
There are many states that have been observed so far in the light meson sector, 
whereas quite some of the observations still need confirmation. Comparing the experimental situation to the 
predictions based on the naive quark model, quite some agreement is found, however, not all the predictions 
are perfectly consistent with experiment: Some of the predicted states have not yet been observed, also 
there are observed states that are not expected, see {\it e.g.}~\cite{PDG}. Given the relatively large 
widths of many of these states ($\sim$100\,MeV/$c^2$) together with the high level density of the mesons 
with (u,d,s) constituent quarks, there are quite some overlaps and mixing, complicating the analyses.

Of particular interest are exotic mesons, not fitting the constituent quark model. QCD allows for and 
predicts such states, like glue-balls, hybrids or tetraquarks according to various models.
Mesons of same spin-parity quantum numbers mix, and the unambiguous observation is thus difficult.
So-called spin-exotic states with $J^{PC}$ quantum numbers not accessible by a simple $q\bar{q}$ 
configuration are on the other hand very promising to search for as they do not mix with ordinary mesons.
The experimental observation of such spin-exotic states would be a fundamental confirmation of QCD, 
for a recent overview, see {\it e.g.}~\cite{MeyerHarlem}.  

The COMPASS data give access to all decay channels spin-exotic hybrid candidates have been 
reported in so far by different experiments, like $\rho(770)\pi$, $\eta(')\pi$, $f_1(1285)\pi$ and 
$b_1(1235)\pi$. The lightest hybrid candidate with exotic $J^{PC}$$=$\,$1^{-+}$, the famous $\pi_1(1600)$, 
has been searched for and is studied in two different decay channels simultaneously, namely in 
$\pi^-\pi^+\pi^-$ and $\pi^-\pi^0\pi^0$ final 
states~\cite{nerling:2009, haas:2011, nerling_meson2012, suhl_hadron2013, florian_PhD_2014, suhl_panic2014} 
produced in diffractive pion dissociation $\pi^-\,p \to (3\pi)^-\,p'$. Since the reconstruction depends on 
different parts of the detector, the observation of a state in both channels provides an independent 
confirmation within the same experiment, the possibility of cross-checks and control of systematics.
\begin{figure}[tp!]
    \begin{center}
      \vspace{-0.5cm}
\resizebox{1.0\columnwidth}{!}{%
  \includegraphics[clip,trim= 10 105 5 125, width=0.75\linewidth, angle=0]{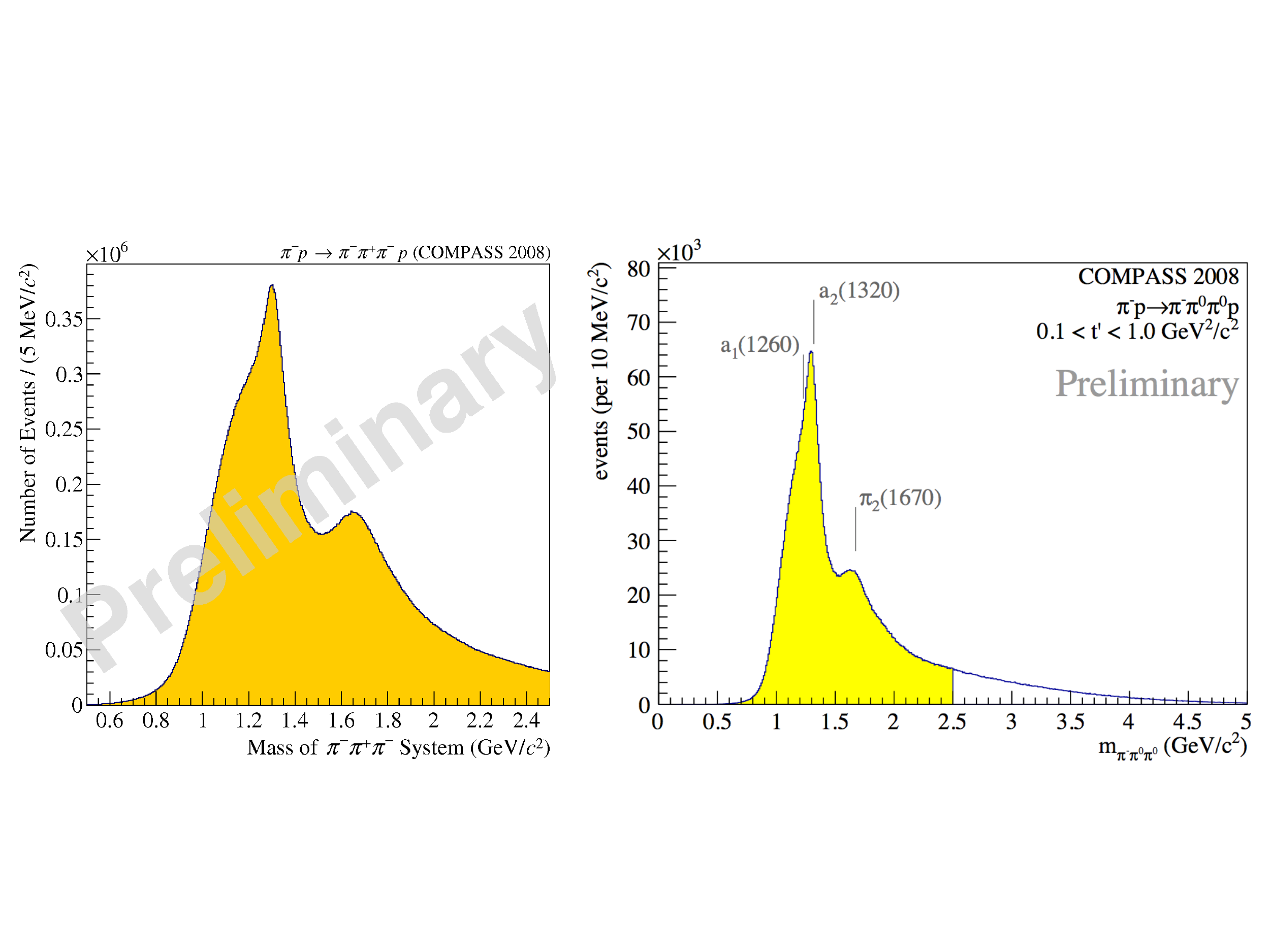} } 
     \vspace{-0.5cm}
     \caption{Invariant mass spectrum of the exclusively produced, outgoing $(3\pi)^-$ system for the 
       charged {\it (left)} and the neutral {\it (right)} decay modes~\cite{florian_PhD_2014,suhl_hadron2013}.}  
        \label{fig:3piMass}
     \end{center}
\end{figure}

The invariant mass of the diffractively produced $(3\pi)^-$ system is shown for the charged and neutral
$\rho\pi$ decay modes in Fig.\,\ref{fig:3piMass} (left) and (right), respectively. 
In the range of four-momentum transfer $0.1$\,(GeV/$c$)$^2$\,$<$\,$t'$\,$<$\,1.0\,(GeV/$c$)$^2$ used for 
PWA, about 50 million and 3.5 million exclusive $(3\pi)^-$ events have been reconstructed from the 2008 
data (of negative pion beam impinging on a proton target) in the mass range of 0.5 and 2.5\,GeV/$c^2$ for 
the charged and the neutral mode, respectively.
By the interaction with the target, the beam pion is excited to an intermediate resonance $X^-$ that 
subsequently decays into a di-pion resonance, the so-called isobar decaying into two pions, and a bachelor 
pion with the relative orbital angular momentum $L$ between them. For the results presented here, the 
$\rho(770)$, $f_0(980)$, $f_2(1270)$, $f_0(1500)$, $\rho_3(1690)$ and a broad $(\pi\pi)_s$ component 
have been used. The set of partial waves contains 87 partial waves up to $L$\,=\,$6$ and an incoherent wave, the so-called 
``flat wave'', having an isotropic angular distribution, representing events of three uncorrelated pions. 

Given the observed dependence on $t'$ and the large statistics, 
the PWA method has been extended with respect to the scheme 
of a two-step PWA as applied previously, for a complete description including all details, 
see ~\cite{compass_PWApaper}. 
The data has been divided into 11 and 8 bins of $t'$ such that about equal statistics is contained in each 
bin, and (as previously) into 20\,MeV/$c^2$ and 40\,MeV/$c^2$ wide $m_{\rm 3\pi}$ bins for the charged and the 
neutral mode data, respectively.
The first step analysis, the so-called mass-independent PWA, 
is carried out independently for each mass bin and independently for the different ranges of $t'$.  
This maximum likelihood (rank-1) fit of the model to the data extracts the production amplitudes, taking 
into account the detector acceptance. 
Finally, the analysis is completed by the second step analysis, the so-called mass-dependent fit.
In this $\chi^2$ fit of Breit-Wigner amplitudes to a subset of the spin density matrix, resonance parameters 
are extracted, for which not only the intensities but also the interferences are taken into account. Also it 
takes into account the observed $t'$ dependencies by performing a simultaneous optimisation of the resonant 
parameters in all $t'$ regions. 

\subsection{Observation of a new axial-vector state  $a_1(1420)$}
\label{subsec-2-1}
\begin{figure}[tp!]
  \begin{minipage}[h]{.32\textwidth}
    \begin{center}
         \vspace{-0.5cm}
\resizebox{1.0\columnwidth}{!}{%
   \includegraphics[clip,trim= 5 5 5 8, width=1.0\linewidth, angle=0]{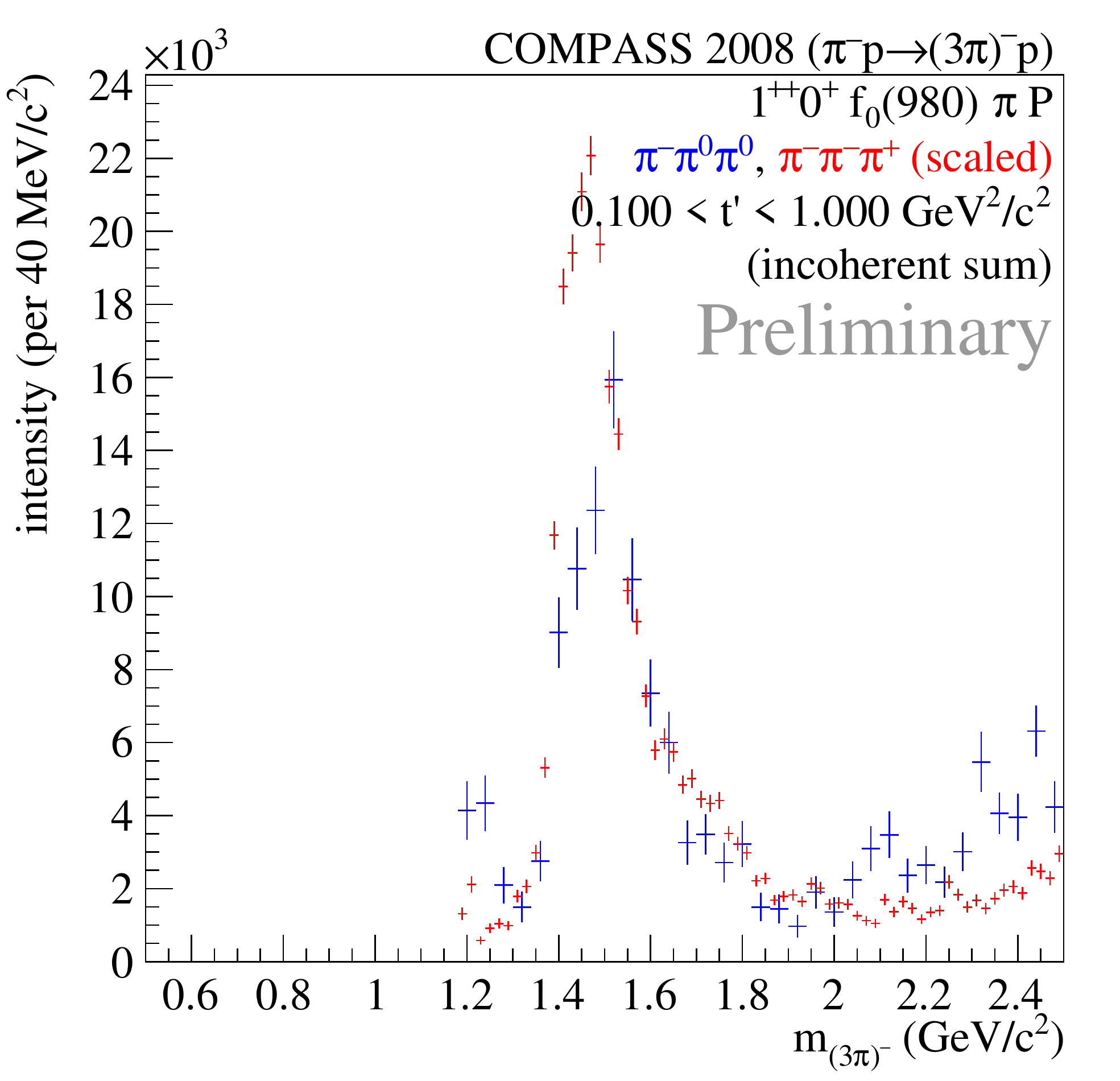} }
    \end{center}
  \end{minipage}
  \hfill
  \begin{minipage}[h]{.32\textwidth}
    \begin{center}
      \vspace{-0.5cm}
\resizebox{1.0\columnwidth}{!}{%
     \includegraphics[clip,trim= 5 5 5 10, width=1.0\linewidth, angle=0]{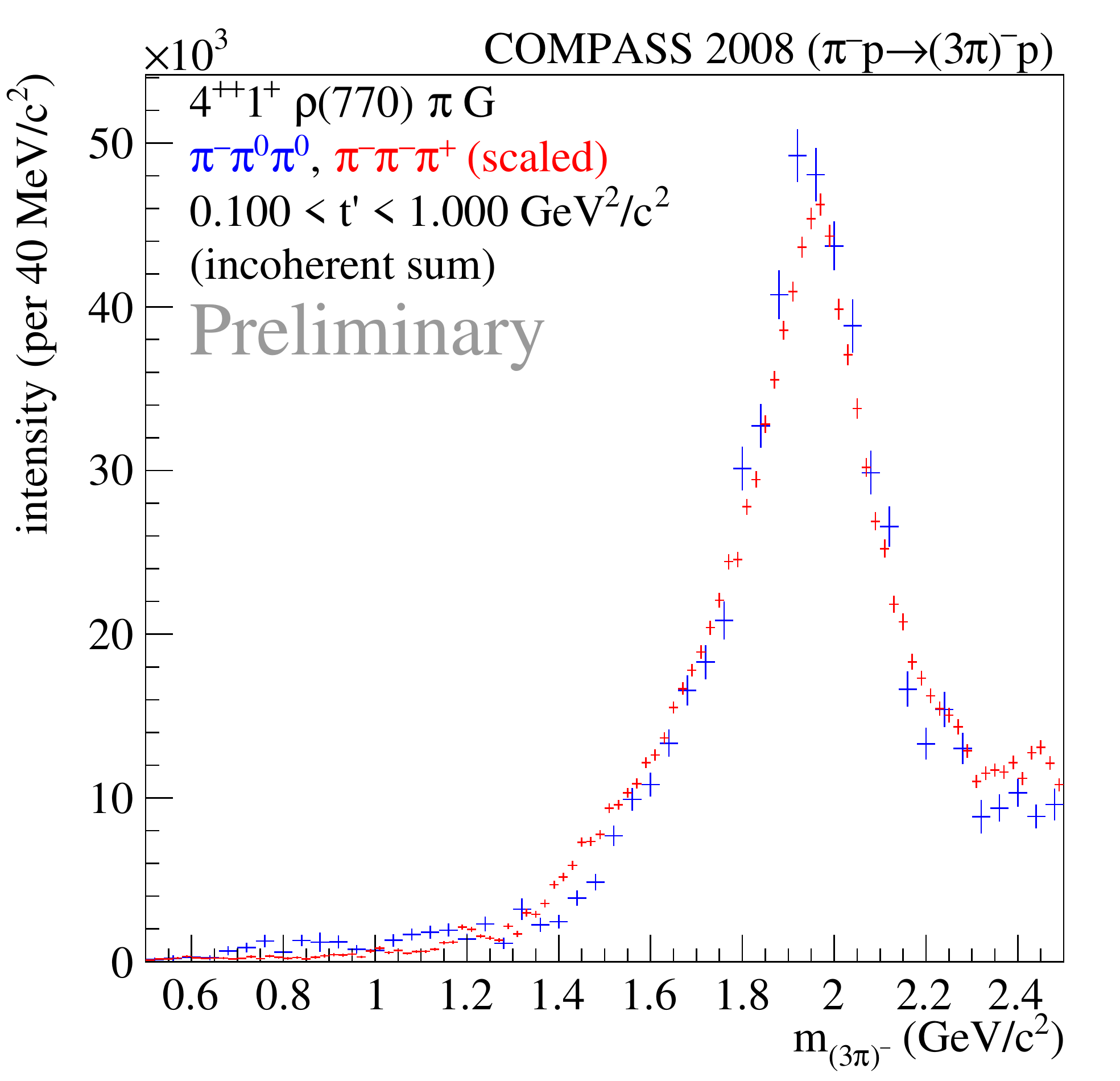} }
    \end{center}
  \end{minipage}
  \begin{minipage}[h]{.32\textwidth}
    \begin{center}
      \vspace{-0.5cm}
\resizebox{1.0\columnwidth}{!}{%
     \includegraphics[clip,trim= 5 5 5 10, width=1.0\linewidth, angle=0]{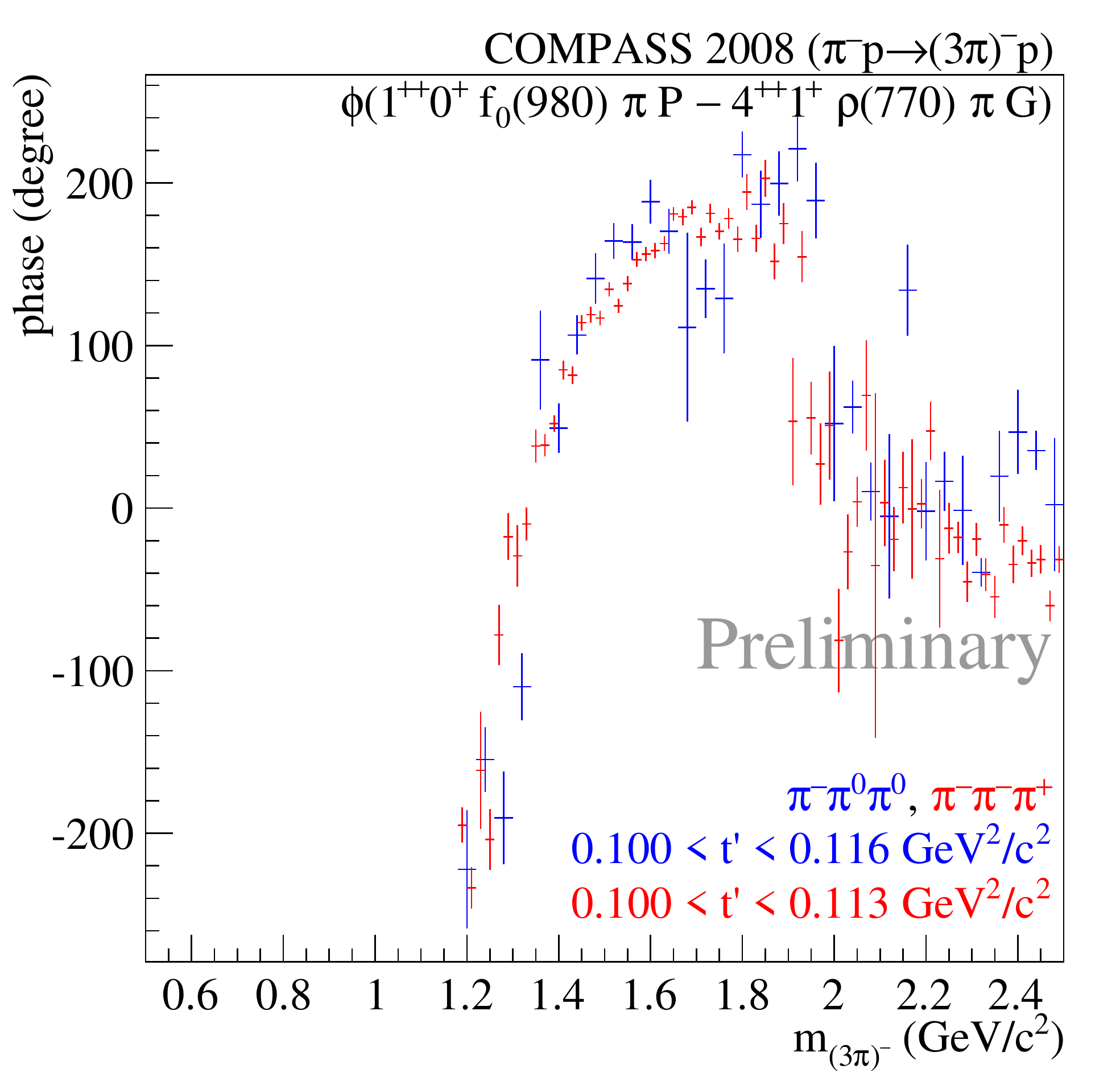}}
    \end{center}
  \end{minipage}
   \begin{minipage}[h]{.32\textwidth}
     \begin{center}
 \resizebox{1.0\columnwidth}{!}{%
  \includegraphics[clip,trim= 5 5 5 10, width=1.0\linewidth, angle=0]{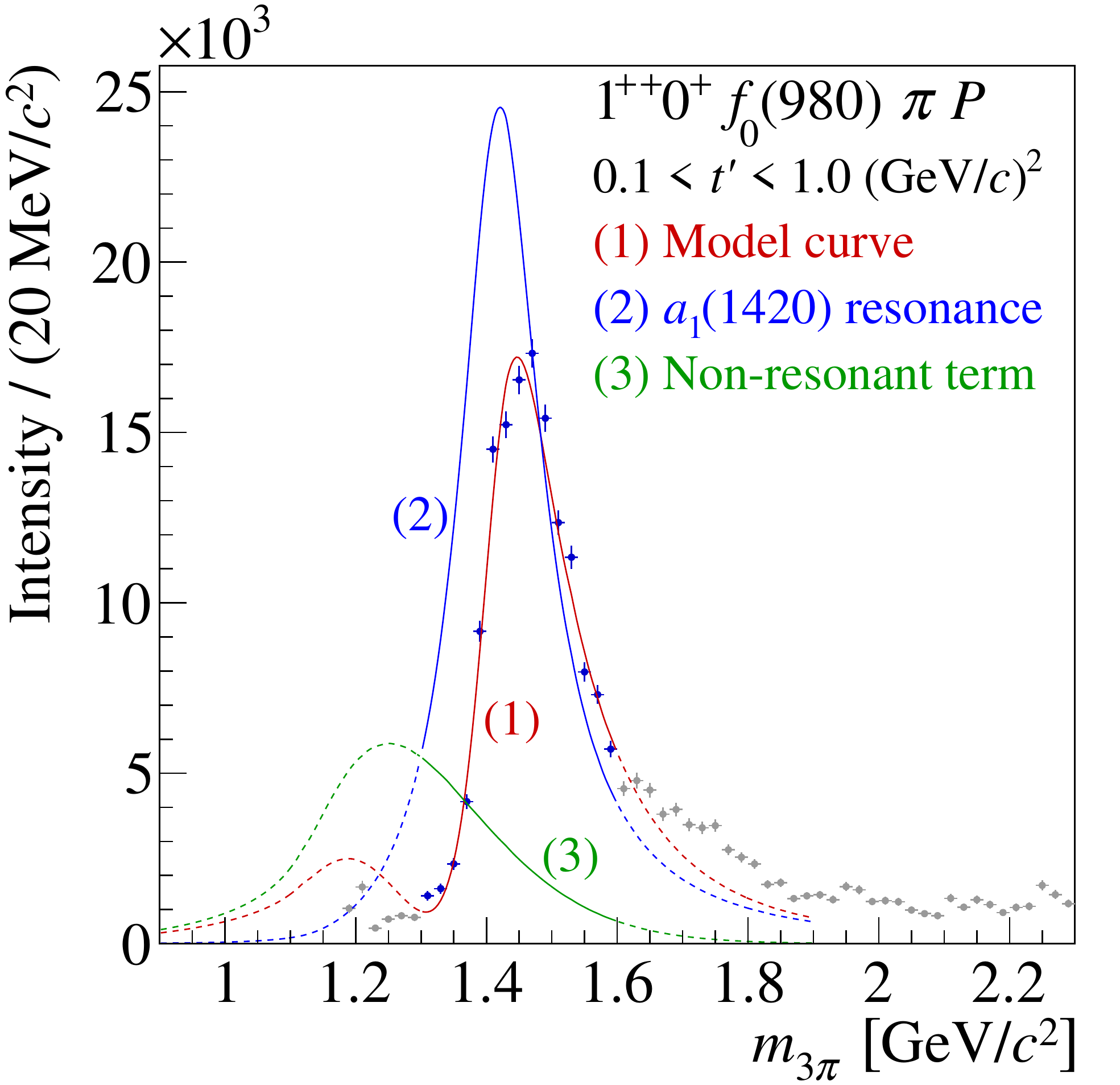} }
     \end{center}
   \end{minipage}
   \hfill
   \begin{minipage}[h]{.32\textwidth}
     \begin{center}
 \resizebox{1.0\columnwidth}{!}{%
     \includegraphics[clip,trim= 5 5 5 10, width=1.0\linewidth, angle=0]{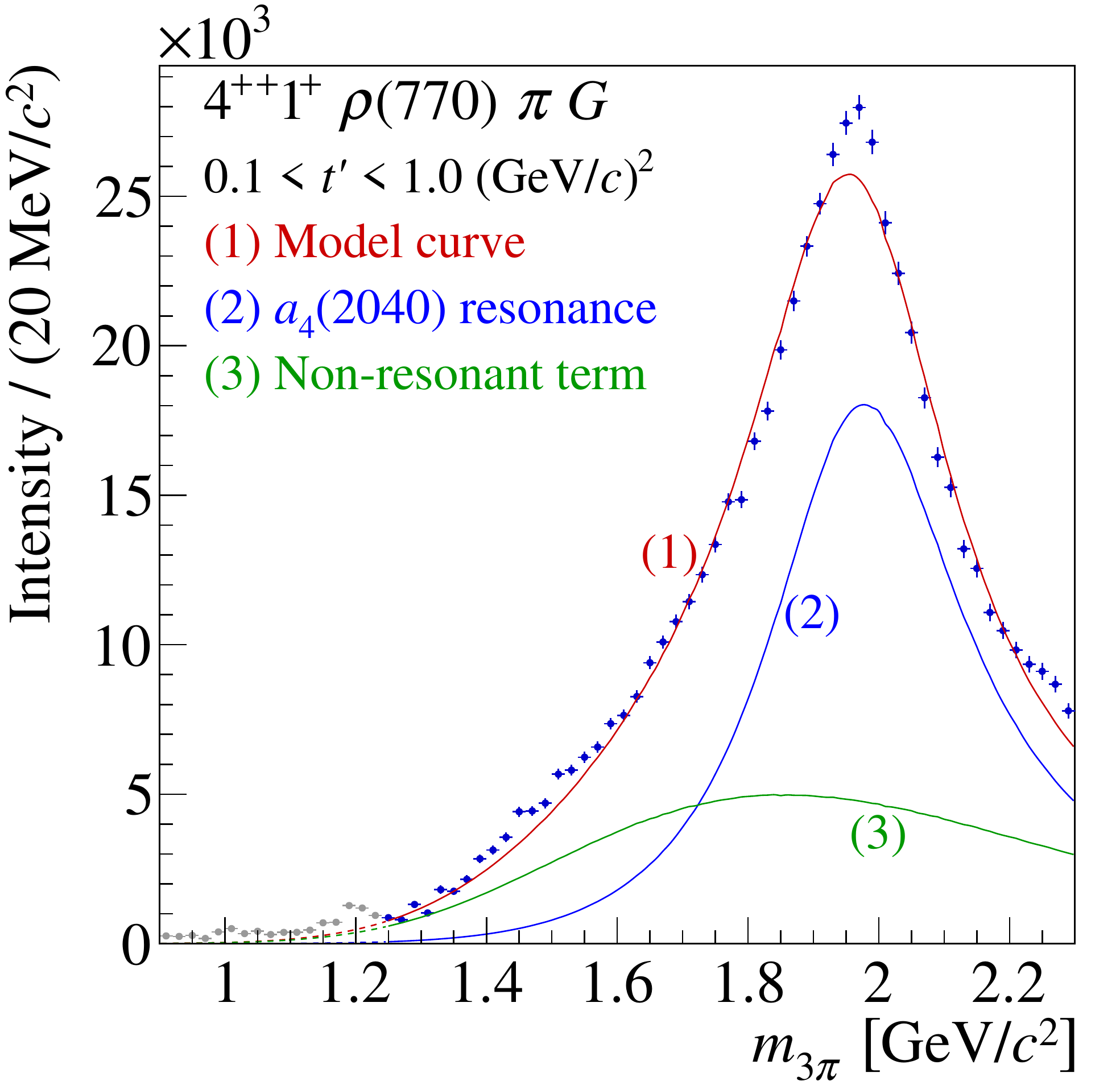}}
     \end{center}
   \end{minipage}
   \begin{minipage}[h]{.32\textwidth}
     \begin{center}
 \resizebox{1.0\columnwidth}{!}{%
   \includegraphics[clip,trim= 5 5 5 15, width=1.0\linewidth, angle=0]{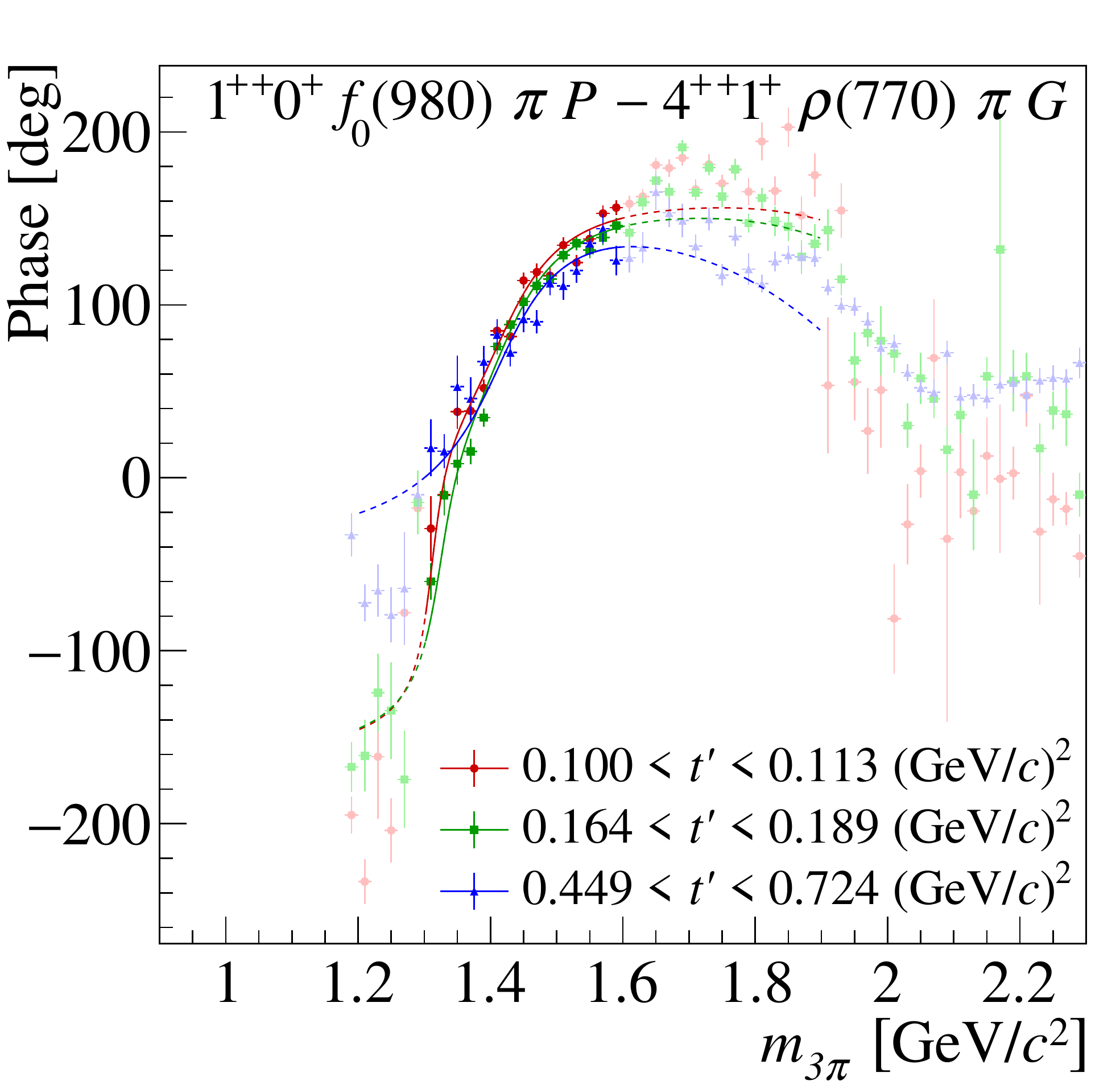} }
     \end{center}
   \end{minipage}
    \begin{center}
      \vspace{-0.5cm}
     \caption{\textit{Top:} First step PWA result, charged versus neutral decay mode~\cite{suhl_hadron2013}: 
       The fitted intensity for the $1^{++}0^+\,f_0(980)\,\pi$\,P-wave shows a narrow structure at about 
       1.4\,GeV/$c^2$, it is similarly observed for both, neutral and charged $(3\pi)^-$ mode {\it (left)}. 
       In the resultant intensity for the $4^{++}1^+\,\rho(770)\,\pi$\,G-wave, the known $a_4(2040)$ is similarly 
       observed {\it (centre)}. The relative phase of the two shows a clean rapid phase motion in the signal 
       region.  
       \textit{Bottom:} Second step PWA result, charged decay mode~\cite{compass_a1}: 
       Complete fit result in terms of the resultant $1^{++}0^+\,f_0(980)\,\pi\,P$ intensity (incoherent sum over t' 
       ranges) {\it (left)}, the first step result is given by the black data points and the resonance model fit 
       overlaid in red consists of the BW describing the narrow structure, that we call the $a_1(1420)$ (blue curve), and the non-resonant 
       background contribution (green curve). The corresponding result is shown for the $a_4(2040)$ observed in 
       the $4^{++}1^+\,\rho(770)\,\pi$\,G-wave {\it (centre)}. The relative phase of the $a_1(1420)$ against the 
       $a_4(2040)$ {\it (right)} is shown for three different t' ranges, consistently showing a clean rapid phase 
       rotation in the signal region.}
       \label{fig:3pi_a1}
     \end{center}
     \vspace{-0.7cm}
\end{figure}

Figure \ref{fig:3pi_a1} (top/left) shows the first step PWA result for one out of the 87 partial waves, namely 
the $1^{++}0^+f_0(980)\pi$\,P-wave, in terms of the fitted intensity for both decay modes. Indeed, the incoherent 
sum over the different t' ranges is shown here, and the charged is normalised to the neutral mode data using the 
integral. Good agreement on the fit result is found for the two channels. For both decay modes, the 
$1^{++}0^+f_0(980)\pi$\,P-wave intensity, containing only a very small fraction of the total observed intensity of 
about 0.25\,\%~\cite{compass_a1}, shows a clear signal slightly above 1.4\,GeV/$c^2$. Rapid phase rotations with 
respect to known resonances, like {\it e.g.} the $a_4(2040)$ (Fig.\,\ref{fig:3pi_a1}, top/centre), are observed in the 
signal region (Fig.\,\ref{fig:3pi_a1}, top/right), as it is expected for a resonance. A phase does not depend 
on the amount of amplitude, and the phase motion is observed independent of $t'$.     

In the second analysis step, a Breit-Wigner resonance model with a minimal set of three waves is used, 
namely the $2^{++}1^+\,\rho(770)\pi\,D$, $4^{++}1^+\,\rho(770)\pi\,G$ and $1^{++}0^+\,f_0(980)\pi\,P$
waves. For the charged mode data, the completed PWA result is show for the $1^{++}0^+\,f_0(980)\pi$\,P-wave 
in Fig.~\ref{fig:3pi_a1} (bottom/left) and for the $4^{++}0^+\,\rho(770)\pi$\,G-wave in 
Fig.~\ref{fig:3pi_a1} (bottom/centre), the phase difference between them is given in 
Fig.~\ref{fig:3pi_a1} (bottom/right). From this fit, we obtain a mass of 
$m\,=\,1414^{+15}_{-13}$\,MeV/$c^2$ and a width of $\Gamma\,=\,153^{+8}_{-23}$\,MeV/$c^2$. 
The interpretation of this new isovector state is still unclear. The properties of the $a_1(1420)$ 
suggest it to be the isospin partner of the $f_1(1420)$, especially supported by the strong coupling 
to the $f_0(980)$ (and not to $\rho$) that is interpretable as a $K\bar{K}$ molecule. The $a_1(1420)$
and the $f_1(1420)$ may possibly be the first observed isospin partners for a $K\bar{K}\pi$ moleculartype 
excitation, which suggests further studies of the $K\bar{K}\pi$ final state. A first, very promising look 
at the $f_1(1420)$ observed decaying to $K\bar{K}\pi$ in the (same) COMPASS 2008 data set has already been 
provided~\cite{nerling_bernhard_2011}, including first very preliminary PWA confirming (due to spin 
assignment) the state in the $K\bar{K}\pi$ invariant mass indeed to be the $f_1(1420)$ and not the 
$\eta(1295)$.

\subsection{Status of the search for the spin-exotic $\pi_1(1600)$ resonance}
\label{subsec-2-2}
In the 2004 pilot run data ($\pi^{-}$ beam, Pb target), a significant $J^{PC}$ spin-exotic signal has been observed 
at $1660$$\pm$$10^{+0}_{-64}$\,MeV/c$^2$. It shows a clean phase motion against well-known resonances and is 
consistent with the disputed $\pi_1(1600)$~\cite{Alekseev:2009a}. 
The high statistics of the 2008 proton target data allows the search for exotic states in different decay modes 
in the same experiment~\cite{nerling:2009}. Employing the same PWA method as in~\cite{Alekseev:2009a}, 
the results obtained for the $(1^{-+})1^{+}\rho^{-}\pi^{0}$ and $(1^{-+})1^{+}\rho^0\pi^{-}$ intensity 
and relative phase are similar and consistent with the previous observations~\cite{haas:2011,nerling_meson2012}. 
Apart of the established resonances $a_1(1260)$, $a_2(1320)$, $\pi_2(1670)$, also $\pi(1800)$ and $a_4(2040)$, 
an exotic signal in the $1^{-+}$ wave at around 1.6\,GeV/$c^2$ has been observed, that shows a clean phase motion with 
respect to well-known resonances. These results are consistently obtained for both $\rho\pi$ decay modes, neutral and charged.
\begin{figure}[tp!]
  \begin{minipage}[h]{.48\textwidth}
    \begin{center}
      \vspace{-0.5cm}
\resizebox{1.0\columnwidth}{!}{%
  \includegraphics[clip,trim= 5 5 5 5, width=1.0\linewidth, angle=0]{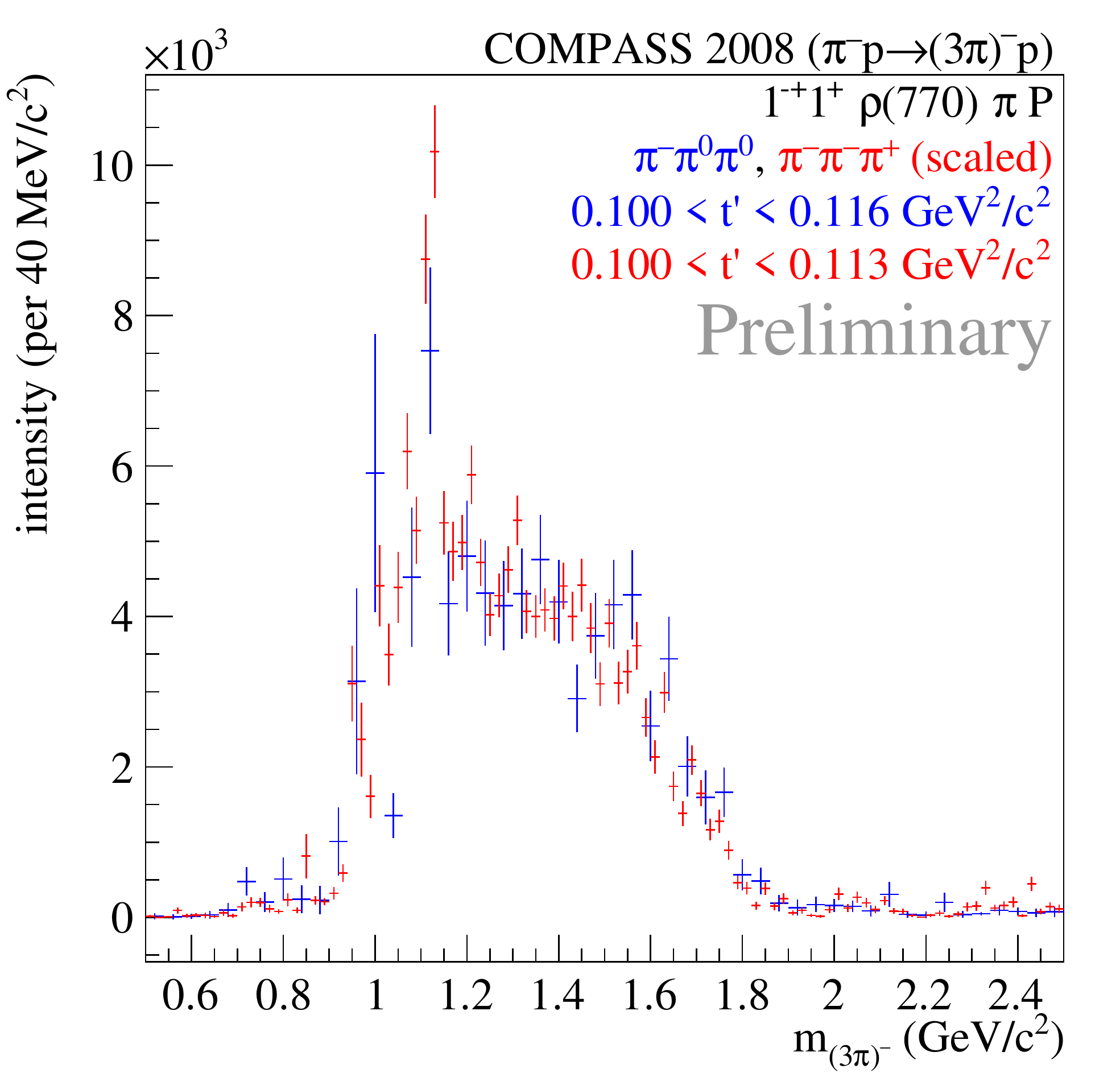} }
    \end{center}
  \end{minipage}
  \hfill
  \begin{minipage}[h]{.48\textwidth}
    \begin{center}
      \vspace{-0.5cm}
\resizebox{1.0\columnwidth}{!}{%
     \includegraphics[clip,trim= 5 5 5 5, width=1.0\linewidth, angle=0]{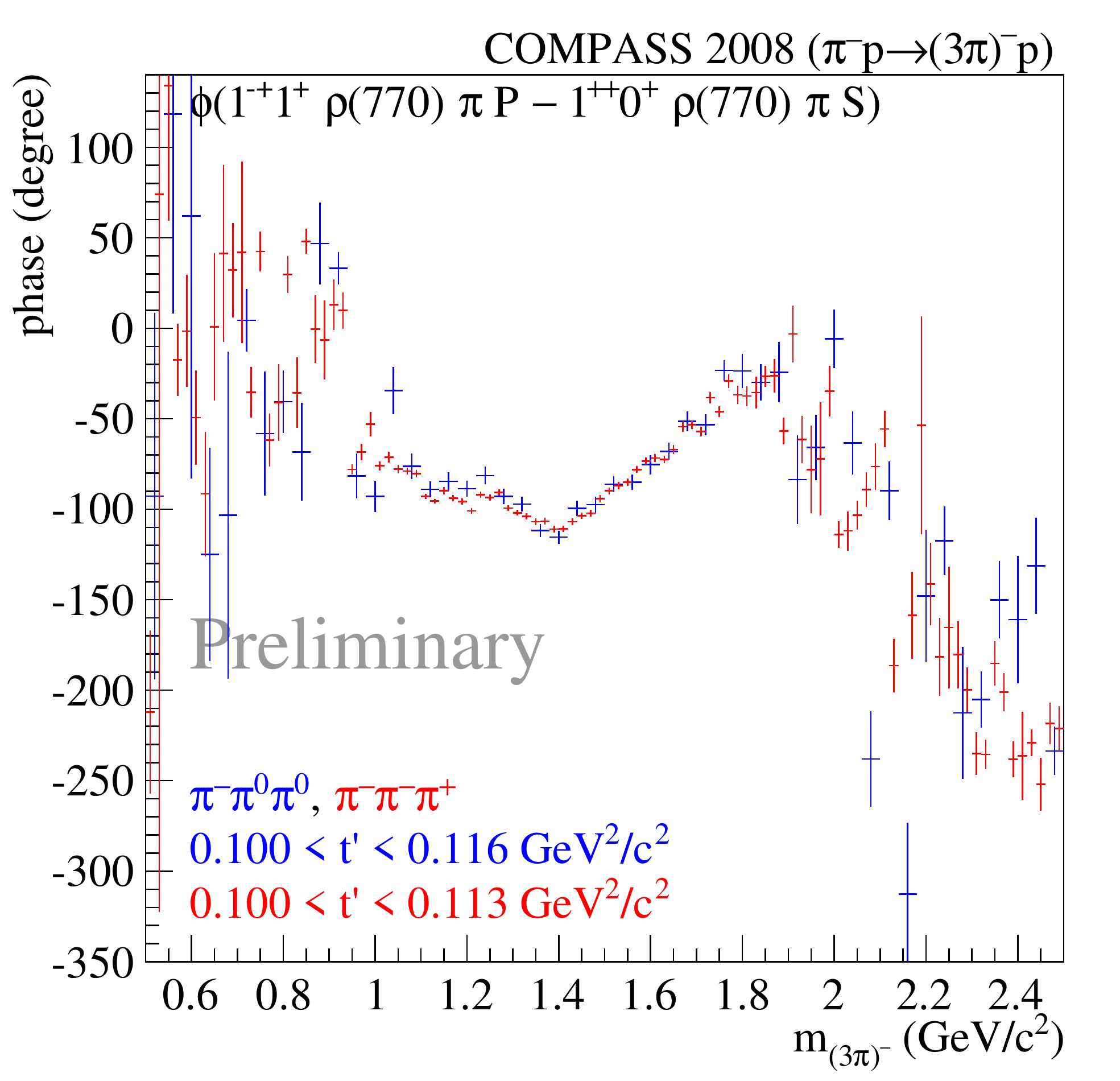} }
    \end{center}
  \end{minipage}
    \begin{center}
     \vspace{-0.3cm}
     \caption{First step PWA result for the exotic $1^{-+}1^+\,\rho(770)\pi$\,P-wave in $(3\pi)^{-}$ final states, charged versus neutral mode data. 
       The resultant fitted intensities are found in good agreement for the two channels {\it (left)}. Shown are the results for 
       the lowest range of four-momentum transfer $t'$, where the signal on top of the broad bump at about 1.6\,GeV$c^2$ has the 
       lowest evidence as compared to the higher t' regions, in which the signal to background ratio improves with increasing $t'$.   
       The relative phases with respect to the $1^{++}0^+\,\rho(770)\pi$\,S-wave, in which the well-known $a_1(1260)$ resonance is 
       observed, is displayed {\it (right)}, showing a clean phase motion in the range of about $1.4$ and $1.8$\,GeV/$c^2$ 
       consistently for the two decay modes.}  
        \label{fig:exoticWave}
     \end{center}
     \vspace{-0.8cm}
\end{figure}
The extension to an additional $t'$ binning and performing the PWA independently also in different 
ranges of $t'$~\cite{compass_PWApaper} allows for disentangling resonant from non-resonant particle 
production ({\it e.g.} dynamically produced components caused by the Deck effect~\cite{Deck:1964}). 

\begin{figure}[tp!]
    \begin{center}
      \vspace{-0.5cm}
\resizebox{1.0\columnwidth}{!}{%
  \includegraphics[clip,trim= 15 125 5 90, width=0.75\linewidth, angle=0]{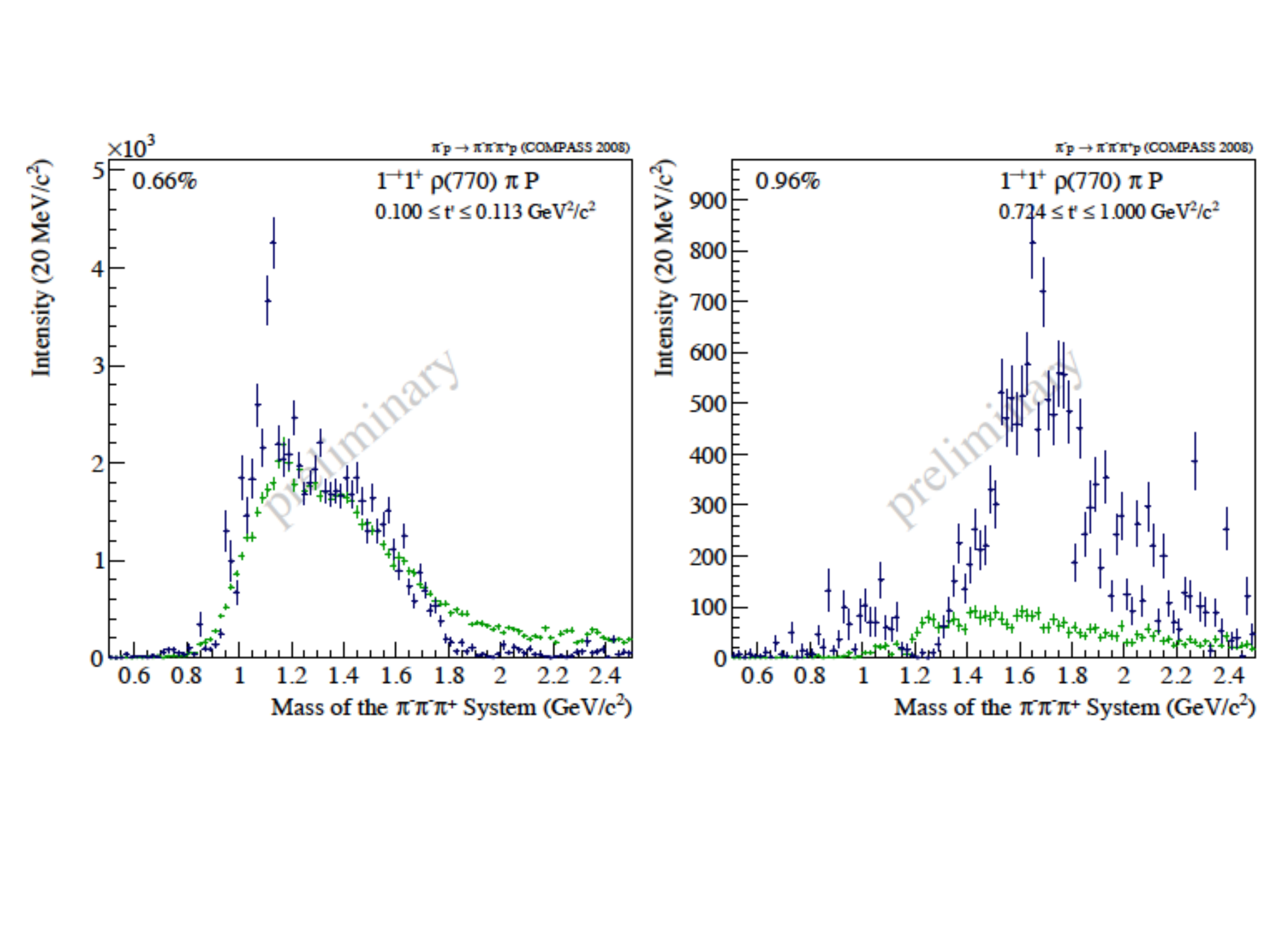} } 
     \vspace{-0.7cm}
     \caption{First step PWA result for the intensity of the exotic $1^{-+}1^+\,\rho(770)\,\pi$\,P-wave for the charged 
       decay mode (blue) overlaid with a projection of a simulated Deck-effect in this wave (green), shown 
       for the range of lowest {\it (left)} and highest {\it (right)} values of $t'$~\cite{suhl_panic2014}.}  
        \label{fig:Deck}
     \end{center}
\end{figure}

A dependency of the exotic $1^{-+}1^+\,\rho(770)\pi\,P$ wave intensity on the squared four-momentum $t'$ is observed. 
Whereas a significant signal is observed at around 1.6\,GeV/$c^2$ on top of a moderate background at larger values of 
$t'$, the signal to background ratio decreases with decreasing $t'$. The fitted intensity of the exotic 
$1^{-+}1^+\,\rho(770)\pi\,P$ wave is shown for the range of smallest $t'$ for the charged and neutral decay modes in 
Fig.\,\ref{fig:exoticWave} (left). Apart from the good agreement between the two channels, we observe mainly a broad 
bump. Even though there is hardly evidence for a signal at about 1.6\,GeV/$c^2$ in this plot, {\it i.e.} for this $t'$ 
range of $0.100$\,$<$\,$t'$\,$<$\,$0.113$\,GeV/$c^2$, the corresponding relative phase with respect to the 
$1^{++}0^+\,\rho(770)\pi$\,S-wave, in which the well-known $a_1(1260)$ is observed, a clean phase motion is consistently 
observed for the two different decay modes in the range of about $1.4$ and $1.8$\,GeV/$c^2$. This is exactly the mass range 
where a significant exotic $1^{-+}$ signal is observed for ranges of larger $t'$ (see e.g. Fig.\,\ref{fig:Deck}, right).  
The relative phase of two resonances is independent of the given absolute amplitudes, and a consistent phase motion is observed
in all bins of t'. 

The observed dependency of the signal to background ratio on $t'$ could be caused by the interference of a resonance with 
a large non-resonant background contribution, possibly introduced and to be explained by the Deck effect~\cite{Deck:1964}. 
In order to further investigate this explanation, pseudo data were generated according to a dedicated model~\cite{Daum:1981}. 
This data has been analysed as the real data, including the decomposition into partial waves, in order to study the Deck-amplitude 
contribution to the different partial waves included in the PWA. Figure~\ref{fig:Deck} shows the promising outcome of the 
study for the exotic $1^{-+}$ partial wave, exemplary for the ranges of smallest and largest $t'$ for the charged decay mode. 
The Deck amplitude intensities are normalised to the total intensity (summed over all mass and $t'$ bins) in the given wave.
A large part of the exotic $1^{-+}$ wave intensity at small values of $t'$ can be described as a contribution by the 
non-resonant Deck process (Fig.\,\ref{fig:Deck}, left), whereas the contribution at large $t'$ is almost negligible according 
to this ansatz.  

For completeness, further relevant results and ongoing analyses that are not dicussed in detail here are the PWA results of 
diffractive $(\eta\pi)^-$ and $(\eta'\pi)^-$ production~\cite{compass_eta} and preliminary PWA results 
for central production ({\it i.e.} via double pomeron exchange) of $\pi^+\pi^-$ and $K^+K^-$ from the 2009 COMPASS proton 
beam data~\cite{CP}.   
\subsection{Search for the manifestly exotic $Z_c(3900)$ state}
\label{subsec-2-3}
\begin{figure}[tp!]
    \begin{center}
      \vspace{-0.5cm}
\resizebox{1.0\columnwidth}{!}{%
  \includegraphics[clip,trim= 12 172 10 70, width=1.0\linewidth, angle=0]{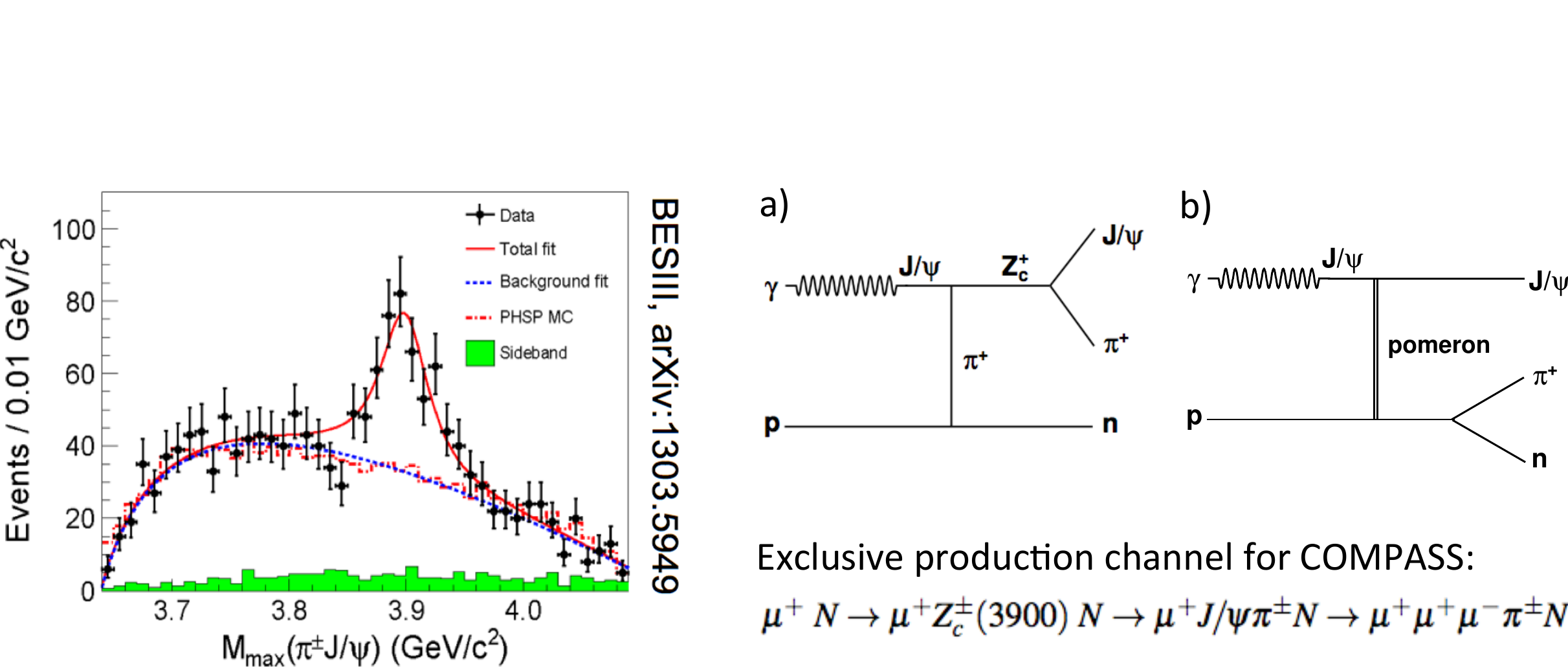} } 
     \vspace{-0.4cm}
     \caption{{\it Left:} The charmonium-like exotic state $Z_c(3900)$ decaying to $J/\psi\pi^\pm$ 
       as observed by BESIII~\cite{bes3Zc}. {\it Right:} In the COMPASS muon data, the $Z_c(3900)$ 
       should be produced via photoproduction according to the process given under a), the main 
       background originates from pomeron exchange reactions, b).}  
        \label{fig:Zc_1}
     \end{center}
\end{figure}

Even though the COMPASS muon beam data is mainly dedicated to study (spin-dependent) 
parton distribution functions, a hadron spectroscopy result has been obtained based 
on the full set of COMPASS data collected with a muon beam between 2002 and 2011, namely 
an upper limit on the photoproduction of the $Z_c(3900)$~\cite{compassZc}. 
The charged charmonium-like exotic state $Z_c(3900)$ was observed decaying to $J/\psi\pi^\pm$ 
by the BESIII~\cite{bes3Zc} Collaboration and confirmed by Belle~\cite{belleZc} in 2013, see 
Fig.\,\ref{fig:Zc_1} (left) for the observation plot, later also the neutral partner 
was confirmed~\cite{bes3Zc_neutral}.

As the photon may act as a $J/\psi$ according to the VDM, the $Z_c(3900)$ should be 
produced via interaction of the incoming photon with the virtual pion from the nucleon 
target according to the diagram in Fig.\,\ref{fig:Zc_1} (right, a), and the cross-section 
for this process should be sizable~\cite{ZcPredict}.

\begin{figure}[tp!]
  \begin{minipage}[h]{.48\textwidth}
    \begin{center}
      \vspace{-0.5cm}
\resizebox{1.0\columnwidth}{!}{%
  \includegraphics[clip,trim= 5 0 5 10, width=1.0\linewidth, angle=0]{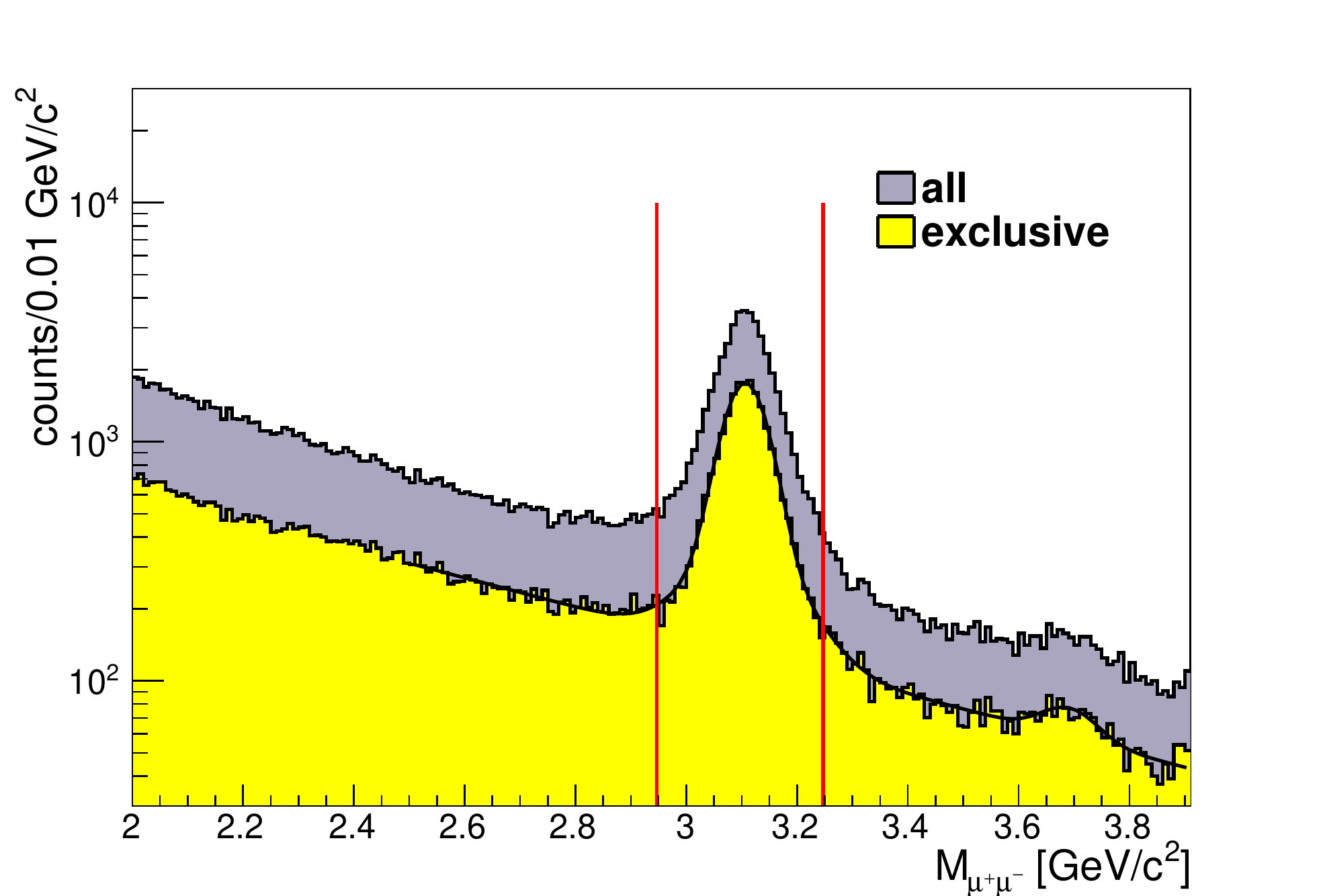} }
    \end{center}
  \end{minipage}
  \hfill
  \begin{minipage}[h]{.48\textwidth}
    \begin{center}
      \vspace{-0.5cm}
\resizebox{1.0\columnwidth}{!}{%
     \includegraphics[clip,trim= 5 0 5 10, width=1.0\linewidth, angle=0]{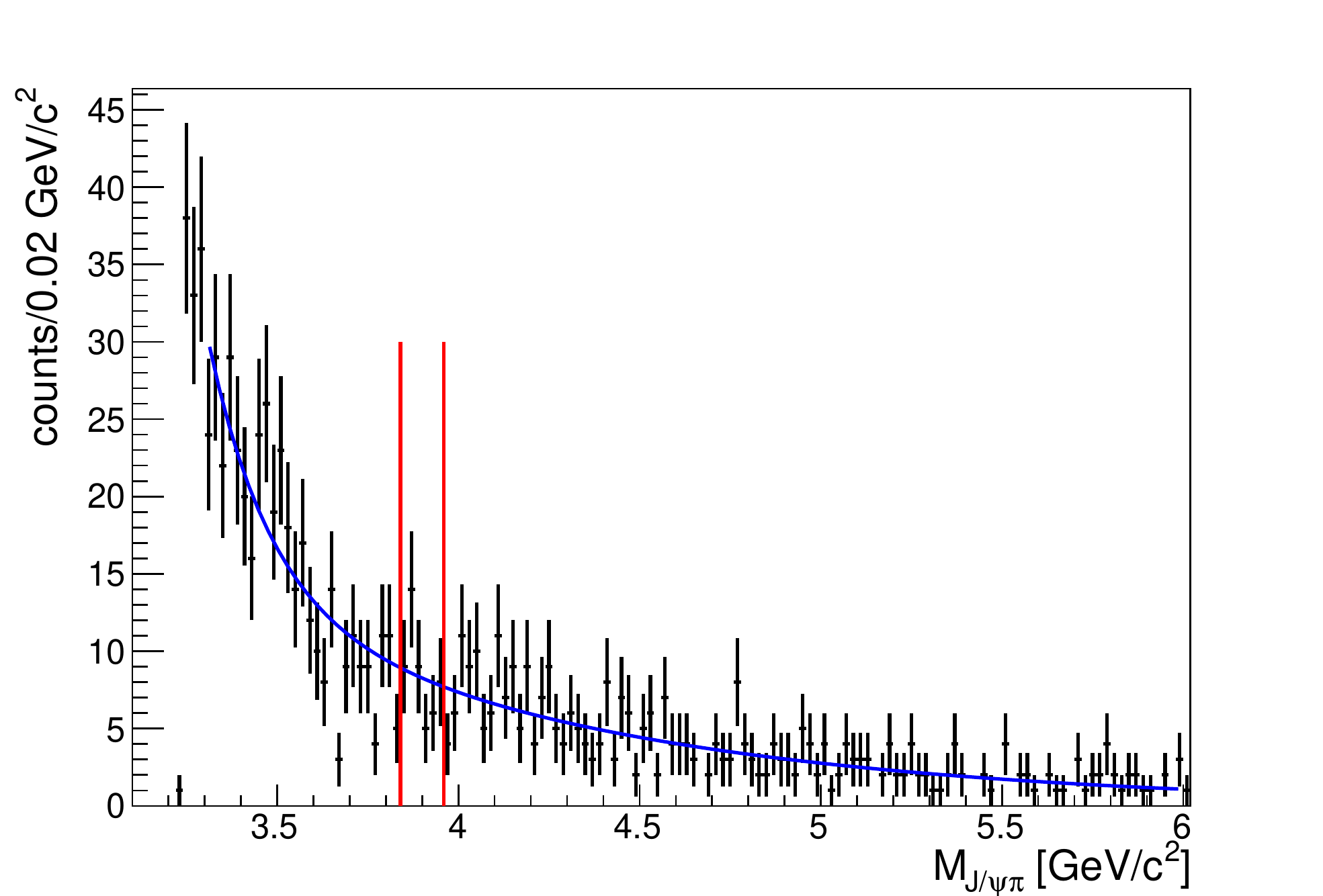} }
    \end{center}
  \end{minipage}
    \begin{center}
     \vspace{-0.3cm}
     \caption{{\it Left:} Reconstructed $\mu^+\mu^-$ invariant mass for all events and for the exclusive sample
       selected, the mass cut applied to identify the $J/\psi$ is indicated.
       {\it Right:} Final plot of the reconstructed $J/\psi\pi$ invariant mass distribution, where the mass range of the 
       expected $Z_c(3900)$ signal is indicated --- no signal is observed~\cite{compassZc}.}  
        \label{fig:Zc_2}
     \end{center}
     \vspace{-0.8cm}
\end{figure}

The reconstructed exclusive production channel is 
$\mu^+ N \to \mu^+ Z^{\pm}_c(3900) N \to \mu^+ J/\psi\pi^\pm N$, with $J/\psi$ decaying to 
$\mu^+\mu^-$, means the final state to detect is $\mu^+\mu^+\mu^-\pi^\pm N$.
The event selection of the exclusively produced $\mu^+J/\psi\pi^\pm$ is straight forward:
A vertex is reconstrcuted having exactly three outgoing muons and a pion, to identify 
the $J/\psi$, a mass cut on the reconstructed $\mu^+\mu^-$ invariant mass is applied as 
indicated in Fig.\,\ref{fig:Zc_2} (left), energy balance is requested and, in order to 
suppress background from pomeron exchange reactions (Fig.\,\ref{fig:Zc_2}, right, b), 
a momentum cut $p_{\pi^\pm}$\,$>$\,$2$\,GeV/c 
for the pions is applied.
In addition, in order to make use of the measurement of the cross-section 
$\sigma(\gamma N$\,$\to$\,$ J/\psi N)$ provided by the NA14 experiment~\cite{NA14} for absolute 
normalisation, the exclusive $\mu^+J/\psi$ sample is selected as well using the same 
selection criteria. The ratio of acceptances for both samples equals in first approximation 
about the acceptance for the additional pion that we know to be $0.5\pm 0.1_{syst}$, 
averaged over all setup and target configuartions.   

The resultant COMPASS plot for the search for the $Z_c(3900)$ is given in Fig.\ref{fig:Zc_2} (right), where the reconstructed 
$J/\psi\pi$ invariant mass is plotted and the expectation region for a signal of the $Z_c(3900)$ is indicated. 
No signal of exclusive photoproduction of the $Z_c^{\pm}(3900)$ state and the decay into $J/\psi\  \pi^{\pm}$ has been found.  
Thus, an upper limit was determined on the product of the cross-section and the relative 
$Z_c^{\pm}(3900)\rightarrow J/\psi  \pi^{\pm}$ decay probability normalised to the cross-section of incoherent exclusive 
photo-production of $J/\psi$ mesons.
We obtain the following result~\cite{compassZc}
\begin{equation}
\label{result}
\frac{BR(Z_c^{\pm}(3900)\rightarrow J/\psi\ \pi^{\pm} )\times \sigma_{\gamma~N\rightarrow Z_c^{\pm}(3900)~ N}}{\sigma_{ \gamma~N \rightarrow J/\psi~N}}\Big{|}_{\langle\sqrt{s_{\gamma N}}\rangle=13.8~GeV} 
< 3.7\times10^{-3}\,.
\end{equation}
Assuming $\sigma_{\gamma~N \rightarrow J/\psi~N}=14.0\pm1.6_{\rm stat.}\pm2.5_{\rm syst.}$~nb as measured by the NA14 Collaboration
for $\sqrt{s}_{\gamma~N}=13.7~GeV$ \cite{NA14}, the result can be presented as
\begin{equation}
\label{result2}
BR(Z_c^{\pm}(3900) \rightarrow J/\psi\ \pi^{\pm} )\times \sigma_{ \gamma~N
  \rightarrow Z_c^{\pm}(3900)~N}\Big{|}_{\langle\sqrt{s_{\gamma~N}}\rangle=13.8~GeV} < 52~\mathrm{pb}.
\end{equation}

This result has been obtained within the framework of the $Z_c$ production mechanism proposed in Ref.~\cite{ZcPredict}. 
Unless the assumptions made therein are wrong, a conclusion is that the regarded decay channel 
$Z_c^{\pm}(3900)\rightarrow J/\psi\ \pi^{\pm}$ can not be the dominant one. This result is 
an important input for the clarification of the nature of this charmonium-like $Z_c^{\pm}(3900)$ state that 
due to the charge and the quark content of $c\bar{c}$ is manifestly an exotic one.

\subsection{Summary and outlook}
\label{subsec-Sum}
The COMPASS experiment at CERN/SPS is well suited for high precision measurements in the fields of 
hadron excitations and spectroscopy, completing together with the hadron structure studies the broad and 
rich COMPASS hadron physics programme.   

Chiral perturbation theory has been confirmed by the COMPASS measurement of the pion polarisability 
($\alpha_\pi - \beta_\pi$) at high precision, and chiral dynamics has been observed (again in agreement 
with ChPT) in the process $\pi\gamma$\,$\to$\,$3\pi$ at low relative momenta. Moreover, radiative 
couplings of meson resonances have been measured at unprecedented precision.    

A possible new isovector resonance with $J^{PC}$\,=\,$1^{++}$, a mass of about 1420\,MeV/$c^2$ and a width 
of about 150 MeV\,$c^2$ has been observed in $f_0(980)\pi P$ decay mode. It shows a rapid phase motion with 
respect to established resonances. The extension to a $t'$-resolved partial-wave analysis provides deeper 
understanding of the exotic signal that we observe at around 1.6\,GeV/$c^2$ in the $1^{-+}1^+\,\rho(770)\pi\,P$ 
wave, showing resonant behaviour. The intensity observed at small momentum transfers $t'$ seems to a large 
extent to be attributed to contributions from the Deck effect, whereas merely a negligible Deck contribution 
seems to play a role at larger values of $t'$. Deeper insights on the disputed $\pi_1(1600)$ hybrid candidate
are not only gained via the two $(3\pi)^-$ decay channels but also in the $\eta\pi$ and $\eta'\pi$ channels.
 
As an outlook, the 2012 Primakoff data set with considerably larger statistics will further improve the pion 
polarisability measurement. Thanks to the extended $x_{\gamma}$ range, also a measurement of ($\alpha_\pi + \beta_\pi$)
will be feasible, also a first measurement of the kaon polarisability is foreseen.      
For spectroscopy, the analysis of $(5\pi)^-$, $(\eta\eta\pi)^-$ and also $(f_1\pi)^-$ will allow for 
access towards higher masses, and complete the list of channels for spin-exotic search.

%
%
%

\end{document}